\documentclass[12pt, preprint]{aastex}
\usepackage{emulateapj5}
\newcommand\asca{{\it ASCA}}
\newcommand\rosat{{\it ROSAT}}
\newcommand\ea{et al.}
\newcommand\hst{{\it HST}}
\newcommand\ha{H$\alpha$}
\newcommand\hii{\ion{H}{2}}
\def\o3{[\ion{O}{3}]}
\def\s2{[\ion{S}{2}]}
\newcommand\kms{\ifmmode{\rm\,km\,s^{-1}}\else{${\rm\,km\,s^{-1}}$}\fi}
\newcommand\psc{\ifmmode{\rm\,cm^{-2}}\else{${\rm\,cm^{-2}}$}\fi}
\slugcomment{To appear in The Astrophysical Journal, August 10, 2001}
\shorttitle{The Obscuring Starburst of NGC 6221}
\shortauthors{Levenson et al.}

\begin{document}
\title{The Obscuring Starburst of NGC 6221 and Implications for the
Hard X-ray Background}
\author{N. A. Levenson\altaffilmark{1}, R. Cid Fernandes, Jr.\altaffilmark{1}, 
K. A. Weaver\altaffilmark{1,2}, T. M. Heckman\altaffilmark{1}, and T. Storchi-Bergmann\altaffilmark{3}}
\email{levenson@pha.jhu.edu, cid@pha.jhu.edu, kweaver@cleo.gsfc.nasa.gov, heckman@pha.jhu.edu, thaisa@if.ufrgs.br} 
\altaffiltext{1}{Department of Physics and Astronomy, Bloomberg Center, Johns Hopkins University, Baltimore, MD 21218}
\altaffiltext{2}{Laboratory for High Energy Astrophysics, Code 662, NASA/GSFC, Greenbelt, MD 20771}
\altaffiltext{3}{Instituto de F\'isica, UFRGS, CP 15051, CEP 91501-970, 
Porto Alegre, RS, Brazil}

\begin{abstract}
We present NGC 6221 as a case study of ``X-ray loud composite galaxies,'' 
which appear similar to starbursts at optical wavelengths and
resemble traditional active galactic nuclei in X-rays.
The net optical spectrum of NGC 6221 
is properly characterized as a
starburst galaxy, but
in X-rays, NGC 6221 is similar to Seyfert 1 galaxies, exhibiting
a power-law continuum spectrum, a broad Fe K$\alpha$ line, and
continuum variability on timescales of days and years.
High-resolution images 
reveal that the detected active nucleus is relatively weak,
not only at optical, but also at near-infrared wavelengths. 
An obscuring starburst, in which the interstellar gas and dust
associated with the starburst conceal the active nucleus, 
accounts for these peculiar features.
We demonstrate quantitatively that obscuration 
by column density $N_H = 10^{22} \psc$ combined with
relatively weak intrinsic nuclear activity can produce
an optical spectrum characteristic of the surrounding starburst alone.
While optical surveys would not identify 
the active nuclei that make these 
galaxies significant X-ray sources, such galaxies may in fact be
important contributors to the X-ray background.

\end{abstract}

\section{Introduction}
Since the discovery of the cosmic X-ray background (XRB) 
\citep{Gia62}, great progress has been made in understanding its
origin.  Surveys with \rosat{} successfully resolved 70--80\% of the
soft X-ray background at energies 0.5 to 2 keV into discrete sources
\citep{Has98,Schm98}.
Current work with the {\it Chandra X-ray Observatory} demonstrates that
discrete sources account for similar fractions of the harder X-ray
background, at energies up to 10 keV \citep{Gia01}.  
The problem is not completely solved, however.  Although much of the
XRB is made up of what we might classify as the nuclei of otherwise
normal bright galaxies or typical active galactic nuclei (AGNs), a
significant fraction of the discrete sources are not easily
identified.  If they are AGN, they appear to have unusual properties.
\citet{Mus00} find a population of optically faint sources,
which could either be the active nuclei of dust-enshrouded galaxies or
the first quasars at very high redshifts.  
The obscuration of some AGNs is so strong that
they are not detected in X-rays at all.
If the SCUBA submillimeter sources, for example, 
are not driven entirely by star formation, they may contain
AGNs with Compton-thick tori and little circumnuclear X-ray scattering
\citep{Hor00}.

Among nearby X-ray emitters 
are galaxies that have X-ray characteristics of AGN, but whose optical
spectra are typical of H {\sc ii} regions or starbursts.
These objects may comprise some of the ``unusual'' discrete sources 
of the XRB, seeming innocuous in larger optical
surveys, but generating significant hard X-ray emission.
\citet*{Mor96} dub them ``composite galaxies'' because they
share some characteristics of both starbursts and AGNs, albeit
at different energies.  
We prefer ``X-ray loud composite galaxies,'' however,
to distinguish these objects 
from the larger class of AGN/starburst
composite galaxies that exhibit their dual nature consistently
at all wavelengths 
\citep*[e.g., ][]{LWH01j,Cid01}. 
While physically, members of both classes contain
AGN and starburst components, their observational characteristics
are distinct.

The X-ray loud composite galaxies comprise a small fraction of 
soft X-ray selected
sources.  \citet{Mor96} examined optical spectra of the 
\citet{Bol92} catalog of {\it IRAS}\ sources detected in the \rosat{} 
All-Sky Survey and found 7 of this type among the 210 galaxies they classified,
and they suggest that obscuration, including of the narrow-line region,
may play a role in accounting for their unusual properties.
The optical spectra have the characteristics of starburst
and H {\sc ii} region systems, rather than AGNs, based on 
diagnostic diagrams of their emission line ratios  
\citep[e.g.,][]{Vei87}.  Only on closer examination do the
optical spectra  hint at additional activity, such as broadened
[O {\sc iii}] lines or a weak broad component underneath the Balmer lines.
At X-ray energies, however, the X-ray loud composites appear to be 
typical AGNs,
sharing characteristics of luminosity and variability.

Here we present NGC 6221 as a case study of the 
X-ray loud composite galaxies.  
Advantageously, this galaxy has been observed over a wide 
energy range,
and we utilize 
ground-based optical spectroscopy, 
high-resolution optical and near-infrared images
from the {\it Hubble Space Telescope (HST)},
and X-ray imaging and spectroscopy from \asca{}
and \rosat{} to physically account for the emission sources.
We adopt a distance of 19.8 Mpc to NGC 6221, with
$H_0=75 {\rm \,km\,s^{-1}\,Mpc^{-1}}$ for 
recession velocity $cz = 1482 {\rm \, km\,s^{-1}}$,
so $1\arcsec \equiv 96$ pc.
Although NGC 6221 has been classified as a Seyfert 2,
its optical spectrum is more characteristic
of a starburst galaxy. 
The X-ray emission of NGC 6221
certainly qualifies it as an active galaxy, and,
as we argue below, it is more like a type 1,
in fact.
We propose
the ``obscuring starburst'' model to account for these properties 
and demonstrate in general that a starburst can easily hide an AGN.
While previous studies of the starburst-AGN connection have
faced the initial task of finding starbursts in active galaxies
(e.g., \citealt{Hec97,Gon98}; \citealt*{Gon01}), 
the X-ray loud composite galaxies represent the other extreme
composition,
in which the dominant starburst conceals the
weaker AGN at optical wavelengths.

\section{Optical Spectroscopy\label{sec:optspec}}

\subsection{Observations}
A long-slit optical spectrum of NGC 6221 was obtained on the night of
1997 September 24 with the CTIO 4-m telescope using the Cassegrain
Spectrograph and Loral 3K CCD. The 
$1\farcs5$- (144 pc-) wide slit
was oriented along the paralactic angle
at position angle $68^\circ$. The spectral range covered
was 3650--7300\AA, at a resolution of approximately 4\AA. The seeing was
$1\farcs2$. After the long-slit 
spectrum was reduced and flux-calibrated,
one-dimensional spectra were extracted.  We binned together 3 pixels along
the slit, which corresponds to $1\farcs5$ or 144 pc, except the outermost
spectra, centered $5\farcs25$ from the nucleus, in which 6-pixel
($3\arcsec \equiv 288$ pc) regions were combined.

\subsection{The AGN Narrow-Line Region}

The net optical spectrum of NGC 6221 is typical of a reddened starburst
\citep*{Phi79,Mor88,Sto95}. 
The only sign of non-stellar activity at these wavelengths is
an \o3$\lambda$5007 component broader than and blue-shifted with
respect to H$\beta$. This feature, as well as the early detection of
NGC 6221 as an X-ray source \citep{Mar79},  motivated
\citet*{Ver81} to propose a composite Seyfert 2/starburst
scenario, in which a faint AGN narrow-line region (NLR) is 
superposed on a dominant \hii{} region powered by massive stars.
Our spectra confirm the presence of the NLR-like \o3{}
component, as illustrated in Figure~\ref{fig:O3Hb_profiles}, where
we show the \o3{} and H$\beta$ profiles for the nucleus and six
off-nuclear positions along the slit. In the nuclear
($1\farcs5 \times 1\farcs5$) spectrum, the broad
and blue component dominates the \o3{} profile, extending to more
than 1000 \kms, while in H$\beta$, the narrow component dominates. 

The flux ratio, $R\equiv F_{[O III]}/F_{H\beta}$, 
increases from 0.3--0.5 (typical of starburst nuclei)
within 100 \kms{} of the H$\beta$ peak to $R> 2$ (typical of
Seyferts) for $v > 500$ \kms{} toward the blue, in agreement with the results
of \citet{Ver81} and \citet{Pen84}.
We agree with previous studies that the broadened feature likely
originates in an AGN NLR.  The observed high $R$ in
the blue wing of the lines can in principle be achieved in metal poor
starbursts, but this would be inconsistent with the approximately solar
metallicity inferred for NGC 6221 
(\citealt*{Sto94}; see also \citealt{Dur88}).
The width of this
high excitation component as well as its flux ratio
is more typical of AGNs than of starburst galaxies.

\begin{center}
\includegraphics[width=3.1in]{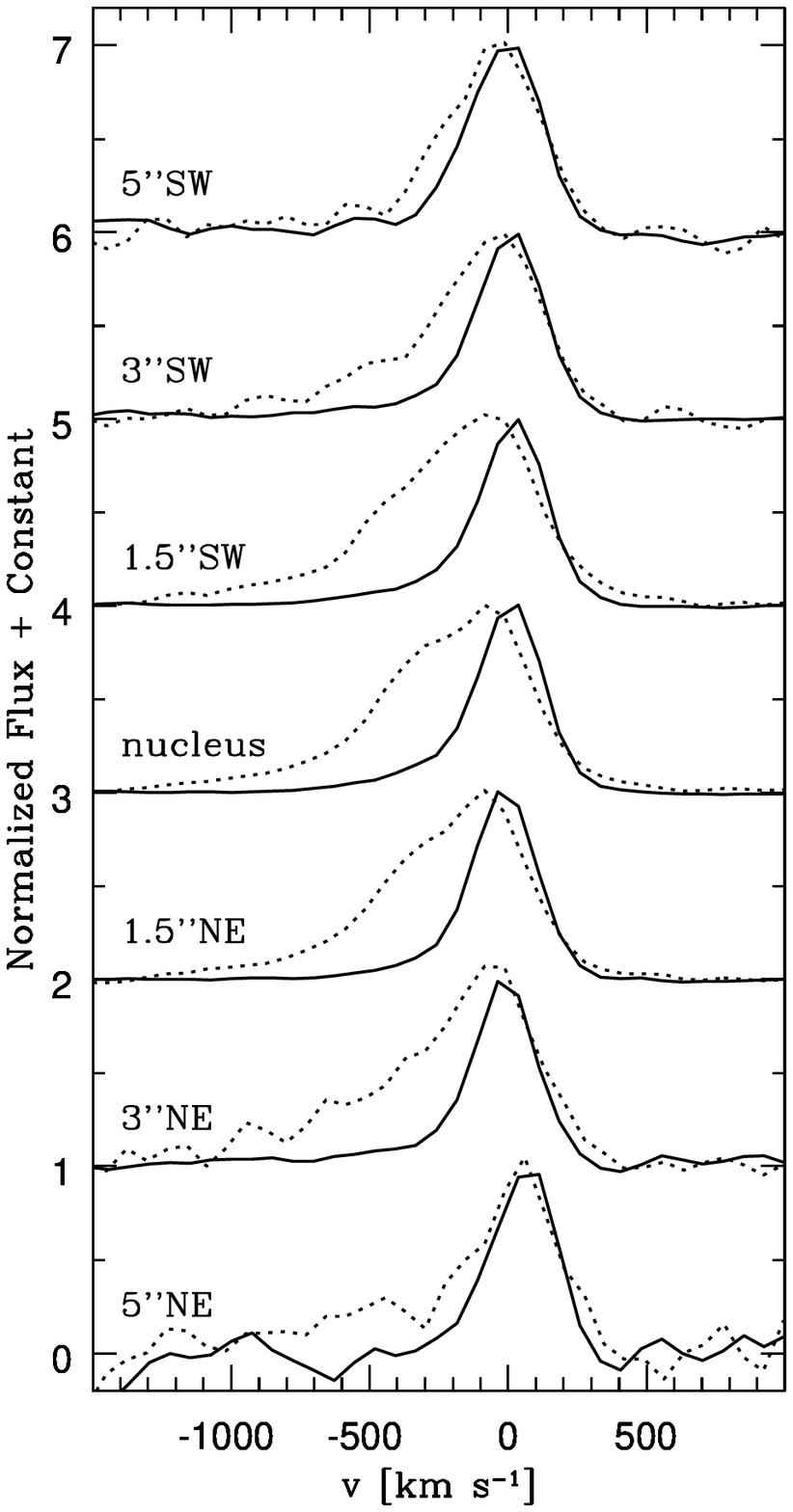}
\end{center}
\vskip -0.1in
\figcaption{H$\beta$ ({\it solid line}) and \o3{} ({\it dotted line}) 
line profiles
across the slit. The five central extractions are $1\farcs5$ wide,
while each of the extreme extractions 
averages the emission
from $3\farcs75$ to $6\farcs75$ from the nucleus. Profiles are
normalized from 0 to 1 and offset by constant values for 
ease of comparison.  The actual \o3/H$\beta$
integrated flux ratio ranges from  0.5 to 0.9.  
The angular scale for NGC 6221 is
$96 {\rm \,pc \,arcsec^{-1}}$, 
so the profiles shown here correspond to
distances up to 500 pc from the nucleus.
\label{fig:O3Hb_profiles}}
\vskip 0.1in

We quantify the characteristics of the two components, modelling the line
profiles with two Gaussians.  We performed a global fit of the H$\beta$
and H$\alpha$ + [\ion{N}{2}] profiles using common kinematic parameters
(centroid and FWHM) for all lines.  The [\ion{S}{2}] lines were not used
in this fit to determine the kinematic parameters but were later deblended
with the two resulting Gaussians.  In all extractions, these combined fits
yield a broad component of FWHM $\sim$ 500--600 \kms{} (after correcting
for instrumental broadening) blue-shifted by 150--250 \kms{} with respect
to a narrower unresolved component (whose intrinsic FWHM $\sim 100$ \kms;
\citealt{Ver81}).  The \o3{} line was modelled separately, with one
component fixed to have the same centroid and FWHM of the narrow Gaussian
identified in the H$\beta$ + H$\alpha$ + [\ion{N}{2}] fits and the other
component unconstrained. This yields a more satisfactory fit to the \o3{}
profile because, even though all lines have the same general structure,
the broad component is much stronger 
in \o3{}.  Fitting the \o3{}
yields a broad component with FWHM $\sim 600$ \kms{} shifted by
approximately $230$ \kms{} with respect to the narrow unresolved
component, which
better represents the kinematic properties of the NLR.  

With this evidence of line width and flux ratios,
we refer to the
broad component as the NLR or AGN component and the narrow one as
the starburst or 
\hii{} component, corresponding to gas photoionized by the AGN and
massive stars, respectively.  
Table 1 lists the fluxes of the main emission
lines for five  positions along the slit.  The percentage of the
flux contained in the NLR component is listed in parentheses
except in the case of [\ion{O}{1}], to which we have not applied the profile
decomposition. 

\begin{center}
\includegraphics[width=3.5in]{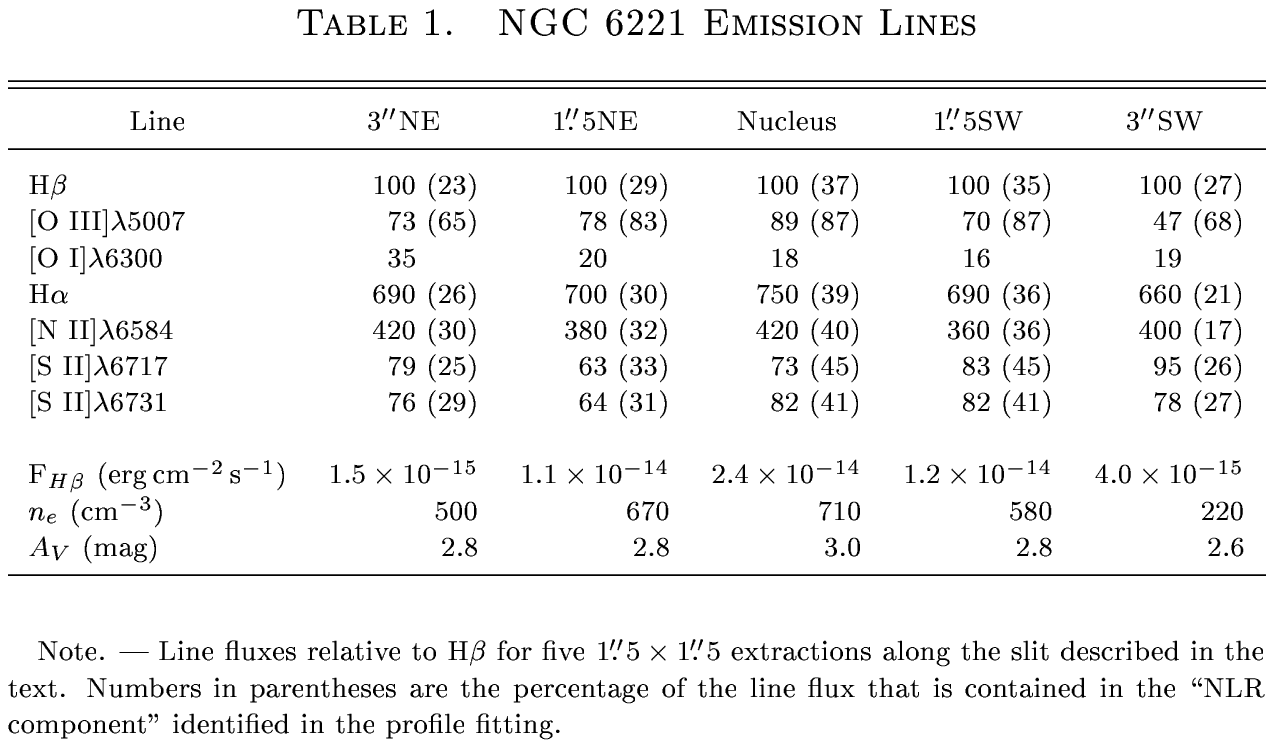}
\end{center}

All the total nuclear emission line ratios are
consistent with a starburst classification on conventional diagnostic
diagrams: 
\o3/H$\beta = 0.89$; [\ion{N}{2}]$\lambda6584$/H$\alpha = 0.56$;
[\ion{O}{1}]$\lambda6300$/H$\alpha = 0.02$; and
[\ion{S}{2}]$\lambda\lambda6717+6731$/H$\alpha = 0.21$.
We note, however, that [\ion{N}{2}]/H$\alpha$ lies
at the boundary between AGN and \hii{}-region-like 
objects when plotted against
\o3/H$\beta$ \citep{Vei87}.  While
the \o3/H$\beta$ ratio for the broad and narrow components yields
NLR and \hii-region-like ratios (2.1 and 0.2 respectively), the
[\ion{N}{2}]/H$\alpha$ and [\ion{S}{2}]/H$\alpha$ ratios are only slightly larger
for the broad component than for the narrow one. Because \o3{} is the
brightest line in NLRs, it is not surprising that our simple
profile decomposition does not disentangle NLR and \hii-like line
ratios for [\ion{N}{2}] and [\ion{S}{2}] as effectively as for \o3.  
Our modelling
of the starburst and NLR components in \S\ref{sec:model} uses only the results
for the H$\beta$ and \o3{} lines, so we have not attempted a more
detailed profile decomposition.
Table 1 also lists the electron density inferred from the
total [\ion{S}{2}] line emission, calculated using the  
IRAF\footnote{IRAF is distributed by the National Optical Astronomy
Observatories, operated by the Association of Universities for
Research in Astronomy, Inc., under cooperative agreement
with the National Science Foundation.} 
task temden.
The density decreases outwards,
from $n_e \sim 700$ cm$^{-3}$ in the nucleus to 100--300 cm$^{-3}$ at 
about 500 pc, in common with other starburst systems \citep{Leh96}.

Besides revealing the NLR component of the AGN,
Figure~\ref{fig:O3Hb_profiles} shows that the region producing this component
is {\em spatially resolved}. The \o3{} asymmetry is easily detected 
in the extractions centered  $3\arcsec$ (corresponding to 290 pc)
away from
the nucleus, and we marginally detect it in the
outer (and noisier) spectra.  
This projected size of 300--500 pc is typical
of observed NLR sizes in Seyferts \citep[e.g.,][]{Schm96}. 
While the blue \o3{} wing gradually fades
farther from the nucleus, its profile increasingly
resembles that of H$\beta$, and
the NLR share of the total line flux decreases (Table 1).

The fact that the NLR feature is seen on both sides of the nucleus
gives an indirect clue of the geometry of the AGN-photoionized gas
in NGC 6221. This symmetry implies that the NLR is not projected as a
one-sided cone.
More importantly, the broad \o3{} component is also
extended along several other position angles 
(H.\ A.\ Fraquelli \& T.\ Storchi-Bergmann, in preparation), 
so the NLR is approximately symmetric in NGC 6221. If
the intrinsic shape of the NLR is roughly conical, then these results
indicate that we see it from within the opening angle of the
cone.  Projected onto the plane of the sky, the NLR thus appears nearly 
circular. This geometry implies that if we could turn off the starburst, NGC
6221 would be classified as a Seyfert 1. 

\subsection{No Sign of a Broad-Line Region\label{subsec:blr}}
There is no indication at all of broad (FWHM $> 1000$ \kms) permitted
emission lines in
NGC 6221, characteristic of Seyfert 1s, 
even at longer wavelengths \citep{Moo88}.
(We stress that the ``broad'' component we refer to above is
{\em not} associated with an AGN Broad-Line Region (BLR), since (1) it is
present in \o3, and (2) it is too narrow for BLR standards.)
Given that the only measurable optical signature of an AGN in NGC 6221 is the
relatively subtle broadened \o3{} feature, we determine
the upper limit on BLR characteristics in the optical spectrum.
By adding a BLR-like (FWHM between 2500 and 7000
\kms) component to the observed H$\beta$ profile, we estimate that any
such BLR component has a maximum equivalent width EW$_{H\beta}^{BLR}$
of 10 \AA; in comparison, the measured equivalent width of H$\beta$ is
21.5 \AA. In terms of flux, this limit corresponds to $F_{H\beta}^{BLR}
< 1.1 \times 10^{-14} {\rm \,erg\,cm^{-2}\,s^{-1}}$, which, correcting
for $A_V =3$ mag (see below), translates into an upper limit
$L_{H\beta}^{BLR} < 1.3 \times 10^{40}{\rm\, erg\,s^{-1}}$ 
on the corresponding intrinsic luminosity.

A Seyfert galaxy without detectable broad optical emission lines
would be classified as a type 2,
as \citet{Ver81} originally proposed for NGC 6221,
but several pieces of evidence 
indicate that its orientation is characteristic of Seyfert 1s.
First, the extended NLR emission noted above suggests 
that we view the nuclear region face-on.
Second, all the X-ray properties, including luminosity,
variability, and emission line attributes, are typical of Seyfert 1s.
We return to this issue in \S\ref{sec:model}, developing a 
physical model of the geometry of NGC 6221 that explains
the absence of broad optical lines despite the orientation.

\begin{figure*}[bth]
\centering
\includegraphics[width=5in]{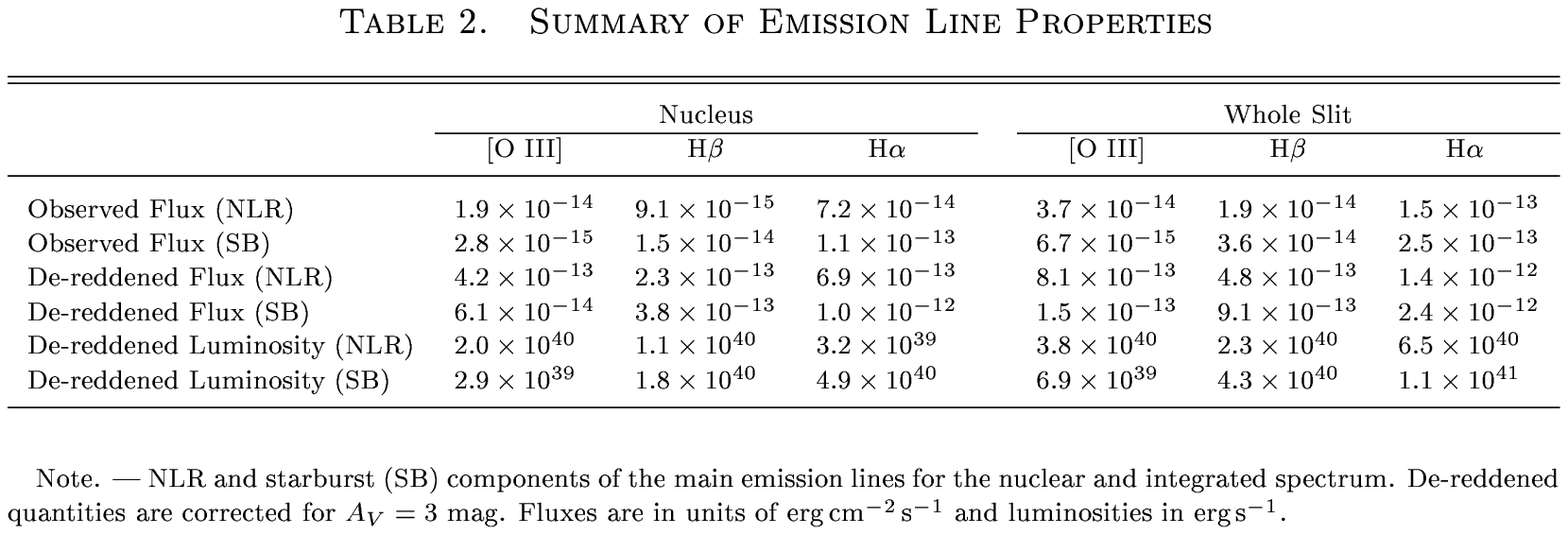}
\end{figure*}

\subsection{Extinction\label{subsec:extinct}}
A distinctive feature of NGC 6221 is its large extinction.  The values
of $A_V$
listed in Table 1
were computed from the measured
H$\alpha$/H$\beta$ ratio using the extinction law of 
\citet*[][with $R_V = 3.1$]{Car89}
and assuming a case B intrinsic
ratio of 2.86 \citep{Ost89}
appropriate to starburst nebulae. In
the nuclear extraction, we measure $A_V = 3.0$ mag,
compatible with the estimate of \citet{Sto94}, and we agree
with their conclusion that the correction for underlying
absorption in the Balmer lines has a negligible effect on the
estimated extinction, so we have not considered it in our analysis.
The H$\alpha$/H$\beta$ ratio for the NLR component is 7.9, somewhat
larger than the 7.2 obtained for the starburst component. This suggests that
that the NLR suffers an additional 0.3 mag extinction with respect to
the starburst lines. Since this is a small difference, we adopt a
common visual extinction of 3 mag for the entire emission line region in NGC
6221. Dust within the Milky Way accounts for $A_V=0.5$ mag 
\citep{Schl98}, 
so the intrinsic extinction in NGC 6221 $A_V = 2.5$.
We summarize the observed
fluxes, de-reddened fluxes, and luminosities of the NLR and starburst
components of several important lines 
in Table~2 for both
the nuclear extraction and the total spectrum integrated from $-6\farcs75$ to
$+6\farcs75$ along the slit. 

\subsection{Properties of the Starburst in NGC 6221}

We characterize the starburst in NGC 6221 using
the Starburst 99 models of \citet{Lei99}. For a constant star
formation rate model with solar metallicity and Salpeter IMF between 1 and
100 M$_\odot$, the star formation rate is related to the H$\alpha$ luminosity
as
SFR $= L_{H\alpha} / (2.98 \times 10^{41} {\rm \,erg\,s^{-1}})
M_\odot {\rm \,yr^{-1}}$. 
Applying this relation to the starburst component in the nuclear spectrum
(Table~2) gives SFR $= 0.17 M_\odot {\rm \,yr^{-1}}$
or $8 \times 10^{-6} M_\odot {\rm \,yr^{-1} \, pc^{-2}}$.
Having such a high central SFR, 
NGC 6221 certainly qualifies 
as a true starburst, rather than a normal galaxy,
compared with the samples of \citet{Ken98sfr}, for example.
The strongest star formation is concentrated within
the central region, extending to a radius of 
approximately 400--500 pc, and it is not confused with
ongoing star formation in the galaxy's disk.

Within the limited circumnuclear region of NGC 6221, however,
the star formation is {\em widespread}, as
the spectral extractions and \hst{} images (\S\ref{sec:hst})
demonstrate.  
This can also 
be seen in Figure~43 of \citet*{Cid98},  where the
equivalent widths of absorption lines and continuum colors are 
plotted as a function of distance from the nucleus. The bluer colors and highly diluted absorption
lines in the extended central region are typical of other starburst and 
Seyfert 2/starburst composite galaxies observed by those authors and others 
\citep{Gon01}.
The widespread star formation is further confirmed 
by \citet{Sto95}, who
measure the H$\alpha$ flux in a
$10\arcsec \times 20\arcsec$ aperture to be 5.4 times larger
than the \ha{} flux we measure in our $1\farcs5 \times
1\farcs5$  nuclear window. 

The far-infrared (FIR) luminosity of NGC 6221 $L_{FIR} = 2.7 \times 10^{10}
L_\odot$. Using the bolometric correction of \citet{Meu99},
this translates into $L_{bol} = 3.7 \times 10^{10} L_\odot$. 
A constant SFR model predicts $L_{bol} \sim 1.1 \times 10^{10}
L_\odot$ per $M_\odot {\rm \,yr^{-1}}$ of star formation, so the implied 
total SFR $= 3.3$ M$_\odot$yr$^{-1}$, 20 times larger than the rate 
inferred for the nuclear ($144 \times 144 {\rm \, pc^{2}}$) region alone.

\section{High-Resolution HST Images\label{sec:hst}}
\hst{}
acquired optical and near-infrared (NIR) images 
of NGC 6221, which we obtained from the \hst{} archive at the Space Telescope
Science Institute.
The Wide-Field Planetary Camera 2 observation provides
spatial resolution of about $0\farcs1 (\approx 10$ pc)
in the $0\farcs046$ pixels of the Planetary Camera (PC), which covered the nuclear region.
The F606W filter was in place, for broad coverage at optical wavelengths; the total
system throughput is centered at wavelength $\lambda_c=6030$\AA, with width 
$\delta\lambda=1500$\AA. 
We use the standard pipeline processing of this single 500 s exposure from 1995 April.
The Near Infrared Camera and Multi Object Spectrometer (NICMOS) camera 2 observed
NGC 6221 for 320 s in 1998 May with the F160W filter ($\lambda_c=1.60\micron, 
\delta\lambda=0.40\micron$). 
The camera's plate scale is $0\farcs075 {\rm\,pixel^{-1}}$, for 
$0\farcs2$ resolution.

We aligned the two images in order to identify the galaxy nucleus
and other sources common to both, although subsequent photometry was
performed on the original calibrated data. 
First we used the nominal
astrometry to perform a two-dimensional polynomial
transformation of the NICMOS image to the 
PC orientation and scale.
Second, we measured the centroids of five point sources common to
both frames and located outside the nuclear region.  We then 
linearly shifted
the transformed NICMOS frame along the rows and columns
by the average measured offsets.

\begin{figure*}[ht]
\centering
\includegraphics[width=6.2in]{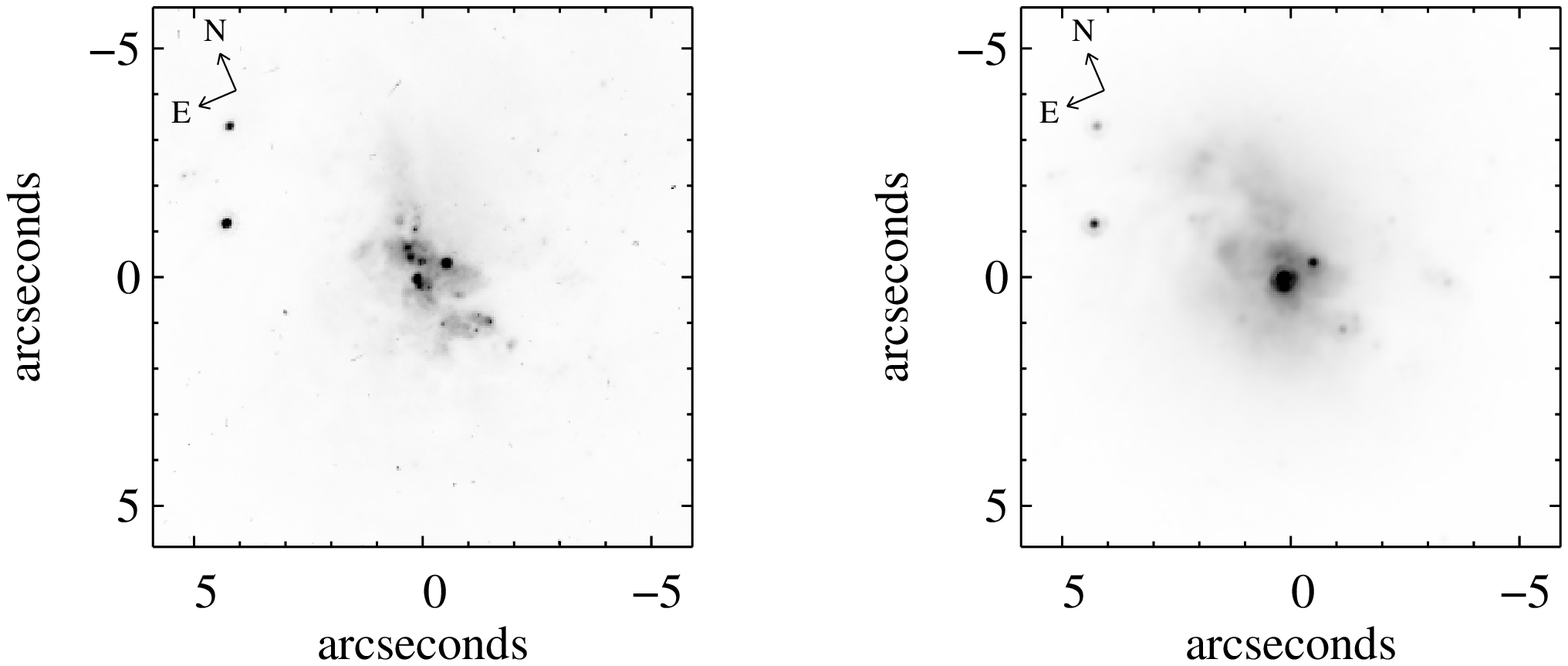}
\caption{\label{fig:hstoptir}
Co-aligned images of the center of NGC 6221 in the optical
({\it left}) and near-infrared ({\it right}).  The brightest
central source detected at 1.6\micron{} is the AGN, although it
is relatively weak at optical wavelengths.
}
\end{figure*}

We display the results in Figure \ref{fig:hstoptir}. 
We identify the brightest central NIR source as the AGN, with
colors and luminosity similar to other active galaxies, 
as we discuss below.
The source is evident in the optical image, but it is not
the brightest one.  In the optical band, the nucleus is relatively weak
compared with the diffuse emission and the other bright sources,
which we identify as stellar clusters.  

We measure the brightness of the nuclear source alone using point
spread function (PSF) subtraction.  Because the PSF varies
significantly across the field of the PC, we obtained a calibration
observation of an unsaturated 
point source located at a similar position in the PC.
We scaled this empirical PSF to match the flux of the AGN within
the central $3\times3$ pixel background-subtracted region,
computing the total flux from the integration of the entire scaled PSF.
In the F606W filter, which is comparable to a $V$ filter, 
the net result is $m_{6060}= 18.1$ mag on the Vega scale, or 
flux density $f_{6060} = 0.19$ mJy. 
Because the NICMOS PSF does not vary significantly across the field, we
use a point source within the field of view to define the PSF.
The PSF fitting yields $m_{1.6}= 13.6$ mag 
on the Vega scale in the F160W filter, 
which is comparable to the standard $H$ band, or $f_{1.6} = 3.8$ mJy.
At 1.6\micron, the PSF subtraction reveals an incomplete ring of 
diffuse emission whose radius is approximately $0\farcs3$ and hints at
a source about $0\farcs1$ away from the nucleus, which is clearly resolved in
the optical image.

The central point source is very red compared with quasars,
and therefore must suffer substantial extinction if it is
the active nucleus.
Specifically, 
we compare with the $V-H$ colors of quasars that \citet{Elv94} observed.  
Although this
sample includes radio-loud as well as radio-quiet objects, it is 
appropriate for comparison because the relative
contribution of the host galaxies is small, so
it yields reasonable measurements of the colors of the nuclei alone.
The nucleus of NGC 6221 equals the median 
$V-H = 2.1$ mag of the sample, 
assuming standard interstellar extinction 
\citep[$A_H/A_V=0.19$; ][]{Car89}
for $A_V=3.0$, the same value
deduced from 
the steep Balmer decrement measured in the optical spectrum
(\S\ref{subsec:extinct}).

The high-resolution images clearly reveal that the nucleus of NGC 6221
is weak with respect to the surrounding starburst, especially at optical
wavelengths.
Through the F606W filter, the nucleus provides only 6\% of the observed flux within the
$1\farcs5 \times 1\farcs5$ aperture of the central region 
in which the central optical spectrum
was extracted (\S\ref{sec:optspec}), as Figure \ref{fig:hstnuc} graphically
illustrates.  
Even at 1.6\micron, where the AGN is relatively bright,
it accounts for only 32\% of the flux within a $1\farcs5 \times 1\farcs5$
region. 

\begin{center}
\includegraphics[width=3.5in]{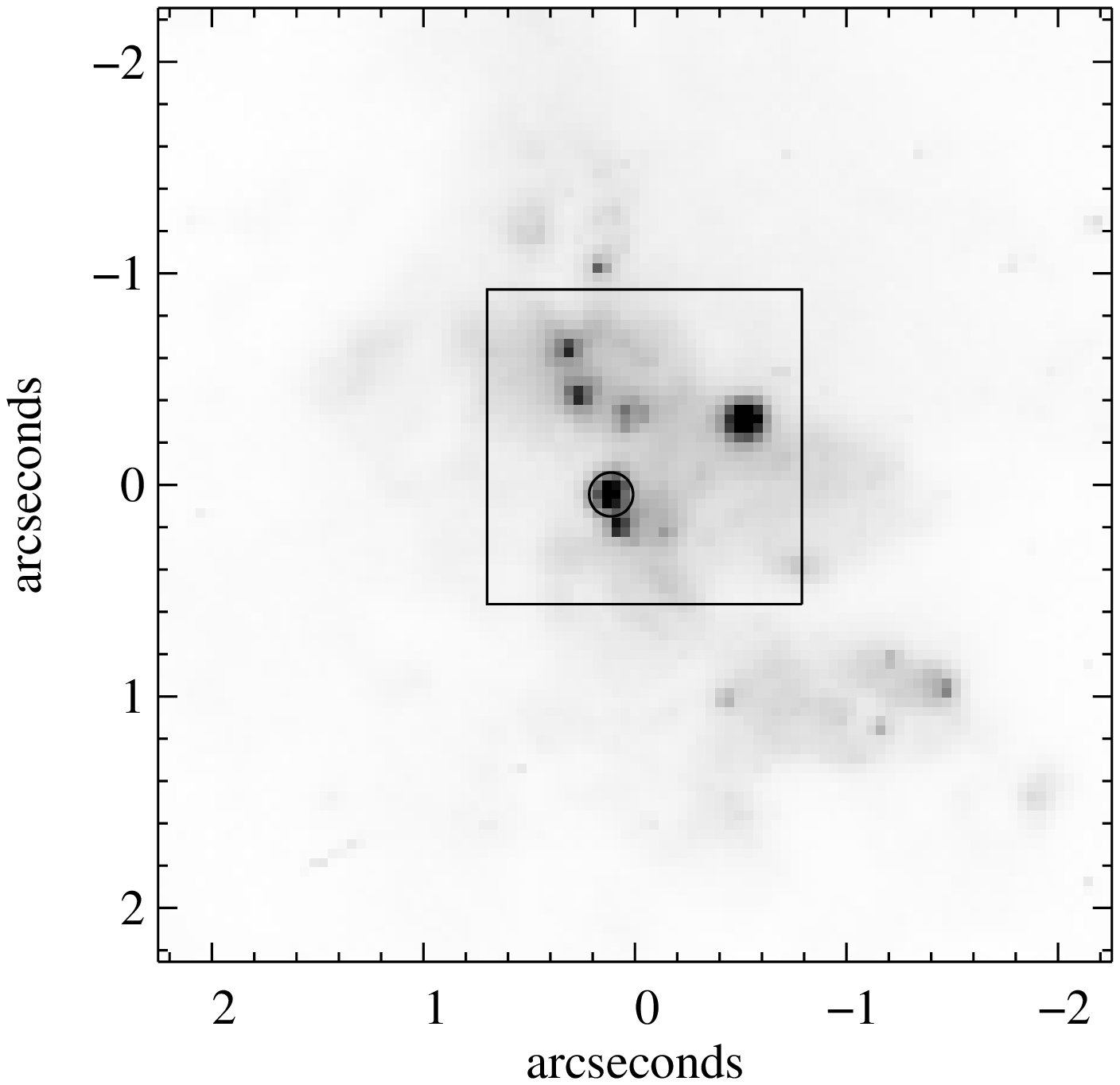}
\end{center}
\vskip -0.1in
\figcaption{\label{fig:hstnuc}
Central region of NGC 6221 observed with the \hst{} PC.
A circle of $0\farcs11$ radius, corresponding to the width of
the point spread function, is drawn around the active nucleus.
The overlaid square is $1\farcs5$ on a side, which corresponds
to the aperture of the central optical spectrum.
}
\vskip 0.1in

\begin{figure*}[bht]
\centering
\includegraphics[width=6.5in]{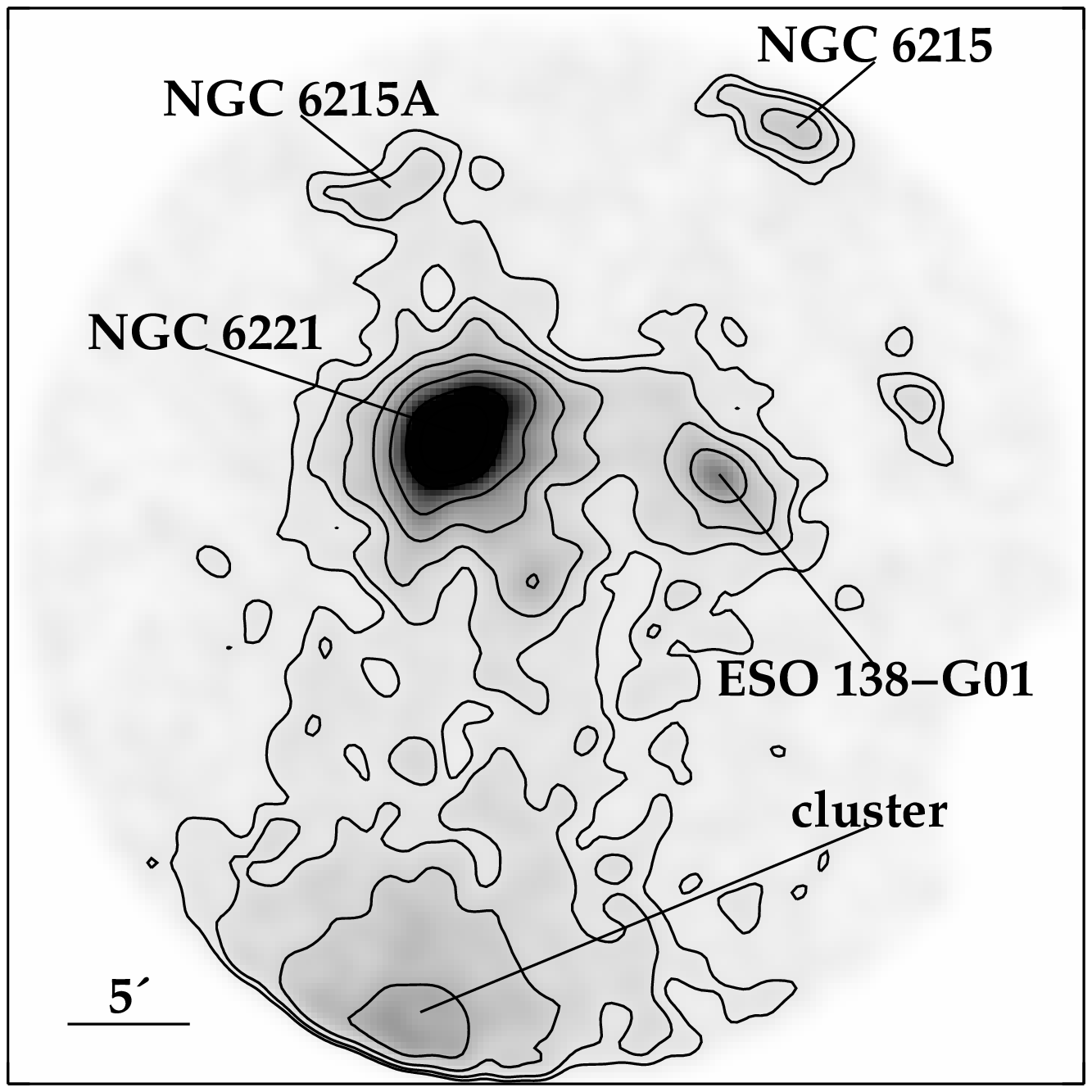}
\caption{\label{fig:ascaimg}
NGC 6221 field observed with \asca{} GIS2.  North
is at the top, and east is toward the left.  The minimum contour
is 2$\sigma$ above the background, and levels are scaled logarithmically
by factors of 2.  The background is not subtracted from the
greyscale image, to illustrate the circular field of view.
}
\end{figure*}

We calculate H$\beta$ flux of the AGN broad-line region
from these measurements of the point source.
For $f \propto \nu^\alpha$, the spectral slope between the $H$ and $V$ bands
$\alpha = -3.1$, so
$f_{4861}=0.096$ mJy.
The equivalent width of H$\beta$ in active galaxies
is approximately 100\AA{} \citep[e.g., ][]{Bor92,Bin93}, nearly all of
which is due to the BLR in type 1 sources.
Thus, we estimate the observed (and extincted)
$F_{H\beta}^{BLR}=1.2 \times10^{-14} {\rm \,erg\,cm^{-2}\,s^{-1}}$
from the combined \hst{} data.
This value is at the detection limit of the optical spectrum
and higher than
the estimates of the intrinsic AGN strength based
on X-ray luminosity we compute below (\S\ref{sec:model}). 
Note that although the extinction in the lines and continuum may
be somewhat different (e.g., \citealt*{Wit92}), in this rough calculation,
we have reasonably assumed that they are equal.

While the evidence for identifying this red central source as the
AGN is indirect, it offers the most plausible account of all the
data.
First, the observed color of this source is typical of 
active nuclei, given the reddening we measure in the spectrum.
Second, the $V$, $H$, and X-ray luminosities simultaneously
match the spectral energy distribution of a typical 
quasar (\S\ref{sec:model}).
Third, the expected line emission of the
point source is consistent with the upper limit on detection
of the BLR.
Finally, none of the other detected sources
share these characteristics, so they are unlikely to be the
active nucleus.
We could fail to detect the AGN entirely, even at 1.6\micron.
Within the bright central region of NGC 6221, we would not detect
a source fainter than $m_{1.6}=18$ on the Vega scale.
In this case, the intrinsic AGN would be extremely weak.
While a faint, buried AGN is certainly possible, it would leave
the problem of explaining the nature of the central source,
with its exceptionally red color 
compared with the surrounding stellar clusters and diffuse emission.
Near-infrared spectroscopy of the red central knot would
clearly test its identification as the AGN.  High spatial resolution
is feasible, and the contrast
between this source and its surroundings is greater at NIR
than optical wavelengths.  The NIR characteristics of
AGN, including high-ionization lines and the BLR component
of hydrogen recombination lines, could be observed.

\section{X-ray Imaging, Spectroscopy, and Variability\label{sec:xray}}

\subsection{Data Reduction}
NGC 6221 was observed with the \rosat{} HRI for 12 ks 
in 1995 September and with \asca{} in
1997 September, resulting in exposure times of 35 ks and 37 ks
in the SIS and GIS detectors, respectively.
The \asca{} data were processed  to exclude times of 
high background rates and to be
appropriate for the SIS 1-CCD observation mode.  The accepted data
were selected to be
outside of and greater than 16 s 
after passages through the South Atlantic Anomaly,
at elevations above the Earth's limb greater than $5^\circ$, 
at times greater than 16 s after a satellite day/night transition, 
and with geomagnetic cutoff rigidity greater than $6 {\rm \,Gev\,c^{-1}}$.
We grouped the spectra into bins having a minimum of 30 counts, so
$\chi^2$ statistics are applicable.  

\subsection{NGC 6221}
The X-ray data are described in detail in \citet*{LWH01s},
and we summarize these results below.
The \asca{} image of NGC 6221 obtained with the GIS2 detector 
(Figure \ref{fig:ascaimg}) 
illustrates the X-ray-emitting environment of this galaxy.
NGC 6221 is in a group.
Extended diffuse emission around
NGC 6221 and ESO 138-G01 implies that these galaxies are
dynamically related, and NGC 6215A may 
also be included in this group.
Observations at 21-cm and a similar recession velocity 
indicate that NGC 6215 
may be interacting with NGC 6221 \citep{Pen84}, 
possibly accounting for its peculiar morphology,
although the X-ray emission from NGC 6215
is separated from the group's emission.
The \asca{} field of view also includes part of the nearby cluster
1RXSJ165259.4-594302, identified in
the \rosat{} All-Sky Survey \citep{Vog99}.

NGC 6221 is
significantly extended in \rosat{} HRI observations,
on a physical scale of at least 5 kpc.  The resolved component
accounts for more than half the soft X-ray emission in this
observation from 1995 September.  By analogy with normal starburst galaxies
lacking active nuclei, this extended component is likely due to
heating of the interstellar medium by stellar winds and supernovae,
possibly including a large-scale outflow where the starburst
``superwind'' has broken out of the galaxy's disk 
\citep{Che85,Hec93}.
The characteristic X-ray spectrum of this diffuse component observed
in pure starburst galaxies and active galaxies that are known to contain
starbursts is soft thermal emission, which is 
due to gas that has been heated
to temperatures of 0.2 to 1.0 keV.

Spectrally, in the
\asca{} observations obtained two years later, 
NGC 6221's AGN characteristics dominate, and the soft
flux increased by a factor of 3.  The total soft (0.5--2 keV) 
X-ray luminosity 
$L_{0.5-2} = 1.4\times 10^{41} {\rm \, erg\,s^{-1}}$,
and the hard (2--10 keV) X-ray luminosity 
$L_{2-10} = 6.6\times 10^{41} {\rm \, erg\,s^{-1}}$,
where both are corrected for Galactic absorption and measured
in the SIS0 detector. 
\citet{LWH01s} model the 0.5--10 keV
spectrum with four components: 
1) the AGN as a power law with photon index $\Gamma =1.9$, 
absorbed by $N_H = 10^{22}{\rm\,cm^{-2}}$;
2) the scattered AGN spectrum, 
with 7\% of the intensity of the intrinsic nucleus;
3) a broad ($\sigma = 0.5$ keV) emission line centered on $E = 6.6$ keV;
and 4) thermal emission with $kT = 1.4$ keV.
Because the X-ray absorption of the intrinsic AGN is small,
even at softer energies (0.5--2 keV),
80\% of the soft X-rays are due to the AGN viewed directly,
and most of the remainder constitute the scattered component.
Only 5\% of the soft emission included in the spectral model is thermal.

The variation in X-ray intensity 
between the HRI and \asca{} 
observations is due to changes in the AGN,
a change of either its
intrinsic luminosity or its obscuring column density.  
The measured flux difference requires either an increase by a 
factor of about 5 in the intrinsic AGN luminosity or a decrease in column
density from $N_H \gtrsim 10^{23}$ at the time of the \rosat{}
observation to 
$N_H = 10^{22} \psc$.
While the relative fraction of
spatially extended X-ray emission, which we associate with
thermal emission \citep{LWH01s,LWH01j}, 
diminished from more than half to only 5\% of the soft flux, its
total flux likely remained constant.

\begin{center}
\includegraphics[width=3.5in]{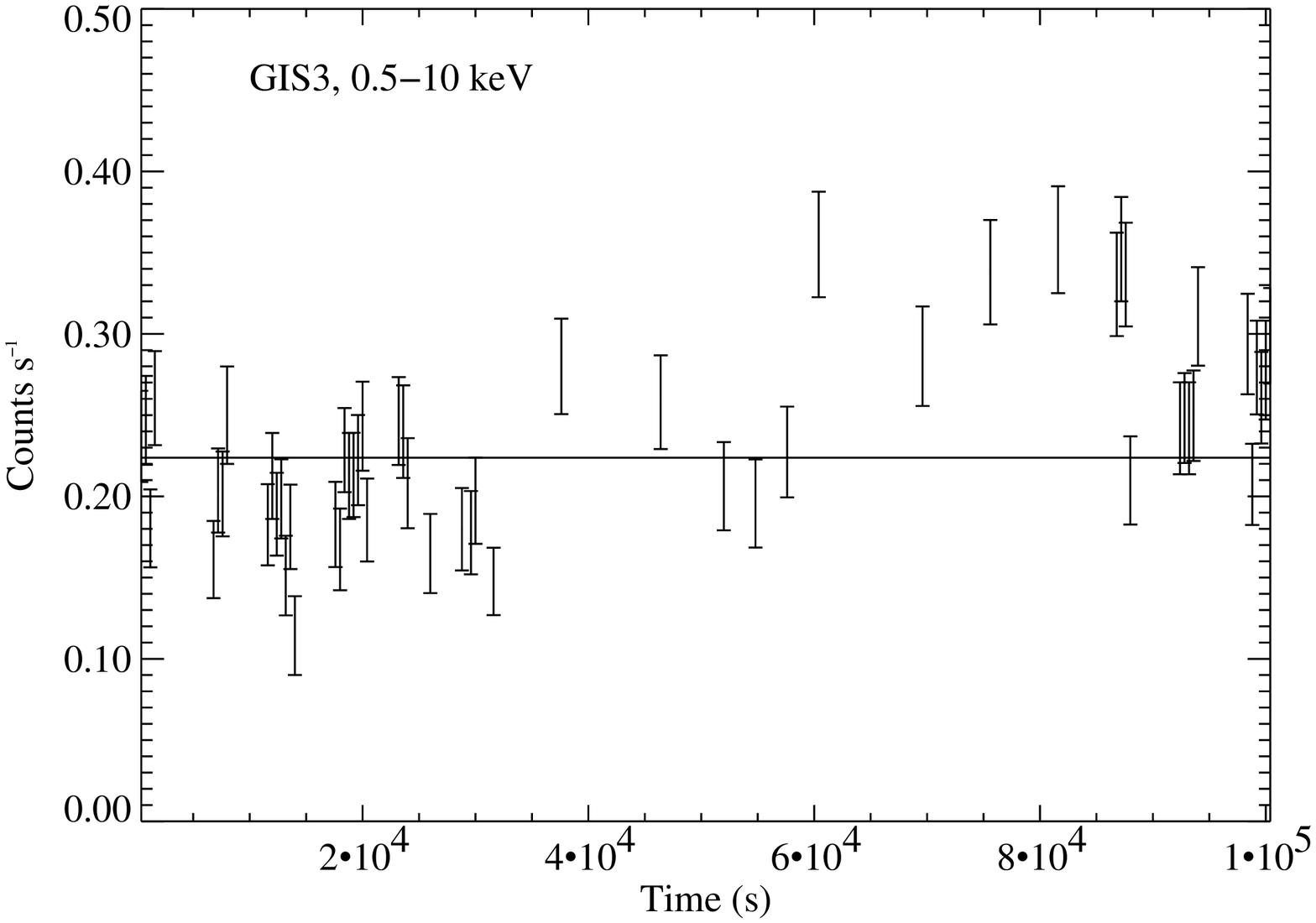}
\includegraphics[width=3.5in]{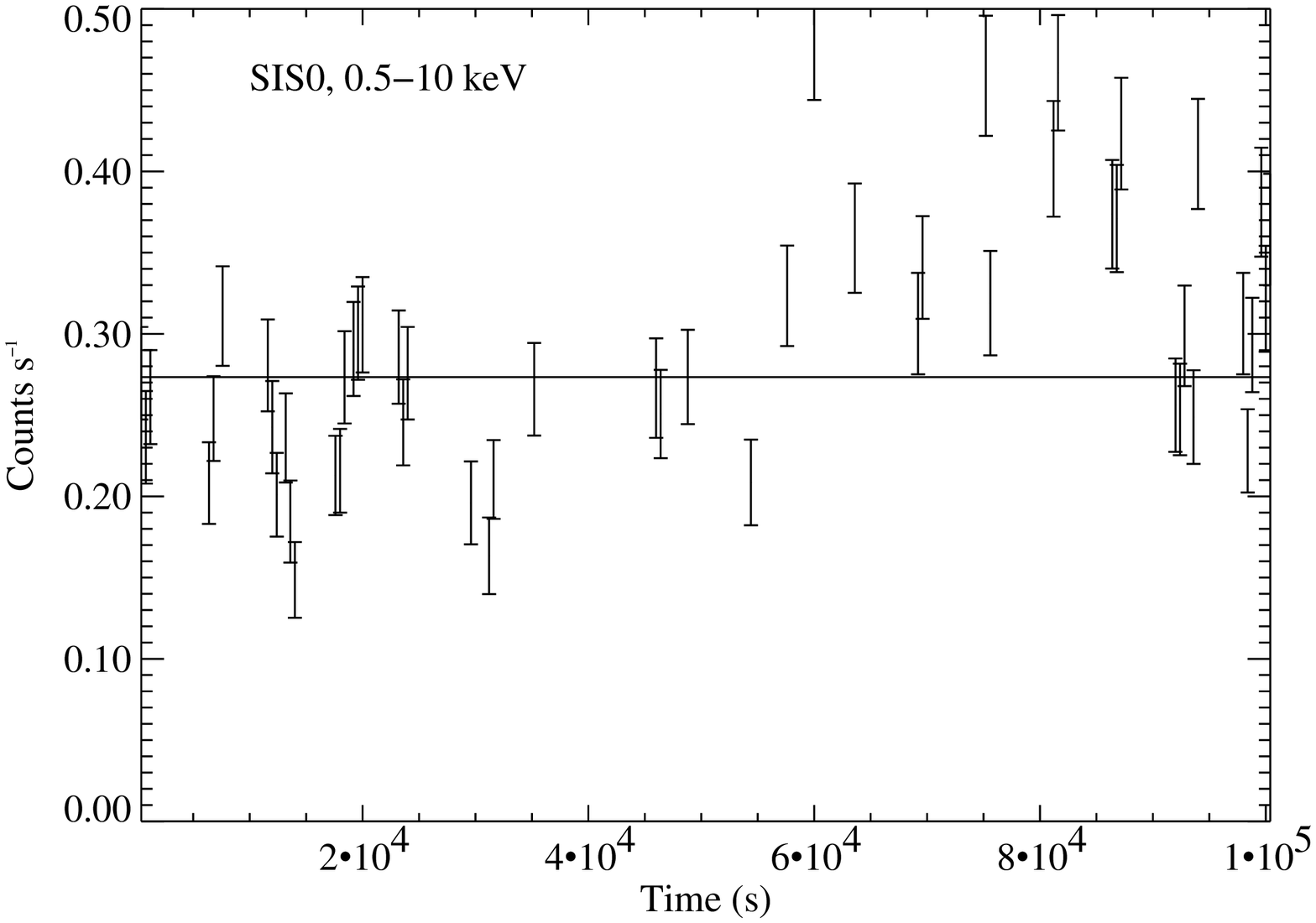}
\end{center}
\vskip -0.1in
\figcaption{\label{fig:lc}
Light curves of NGC 6221 in GIS3 ({\it top}) and SIS0 ({\it bottom})
detectors.  The count rates, extracted in 400-s bins for clarity, have been
background-subtracted.  The best-fitting constant rate for each
case is overlaid.  In neither case does the constant model fit
the data acceptably.
}
\vskip 0.1in

\begin{figure*}
\centering
\includegraphics[width=5.5in]{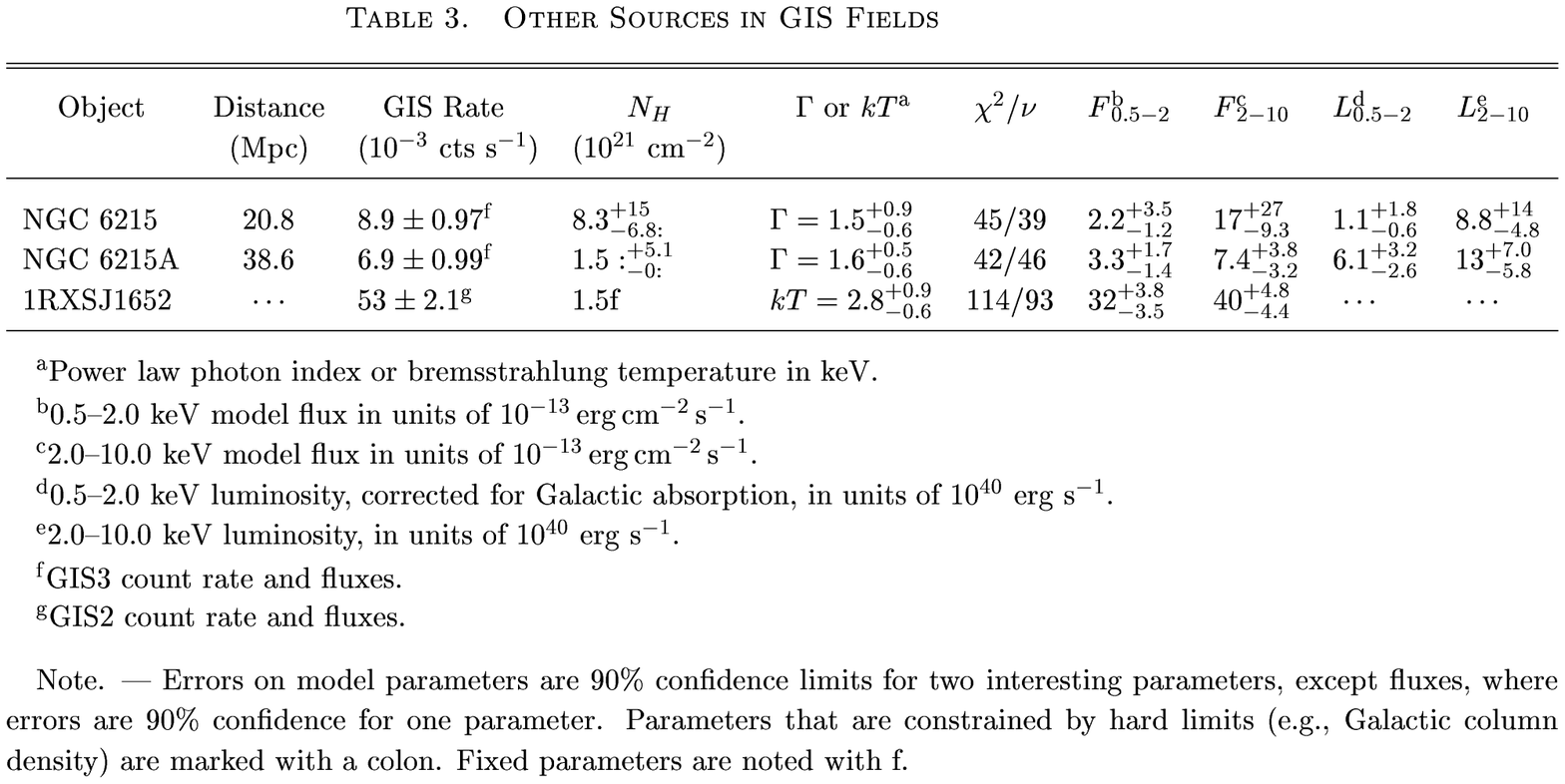}
\end{figure*}

On shorter timescales, within the \asca{} observation, we also
measure significant variability.  
Background-subtracted lightcurves
are shown in Figure \ref{fig:lc} for the GIS3 and SIS0 detectors,
where 400-s bins clearly illustrate the significant variablity.
The best-fitting constant count rate is drawn, 
but this constant flux model fits the data unacceptably in both cases. 
We extracted spectra over limited time intervals to examine
the (earlier) lower flux and (later) higher flux states separately.
With the limited statistics of these shorter exposures, however,
we could not identify significant spectral differences between
the two states.  The same physical model that
fits the total spectrum are also appropriate for the time-constrained
spectra, with measurable variation only in the intensity.
Similarly, the hardness ratio, defined in terms of the hard and
soft fluxes as $(F_{2-10} - F_{0.5-2})/(F_{2-10} + F_{0.5-2})$, 
shows no trend with the variation of total intensity.

The X-ray flux varies by about 50\%  over a
timescale of $5\times10^4$ s.
We quantify the variation in terms of the
``excess variance,'' $\sigma_{rms}^2$, defined by \citet{Nan97a}.
Sampling in 256-s bins in order to compare with published results,
we find $\sigma_{rms}^2 = 0.062\pm 0.014$ for the GIS3 lightcurve
of NGC 6221.
Thus, NGC 6221 is similar to broad-lined Seyfert 1
galaxies in the trend of increasing $\sigma_{rms}^2$ with decreasing
hard X-ray luminosity;
e.g., \citet{Tur99} measure
$\sigma_{rms}^2 \approx 1\times10^{-4}$ at 
$L_{2-10} \approx 10^{45} {\rm \,erg\,s^{-1}}$, rising to 
$\sigma_{rms}^2 \approx 0.05$ for 
$L_{2-10} \approx 10^{41} {\rm \,erg\,s^{-1}}$.

In addition to the unusual variability, 
its Fe line emission makes NGC 6221  unlike most other Seyfert 2s.
It contains a broad  Fe K$\alpha$ line
centered at $E = 6.6$ keV with EW$\sim400$ eV 
and line width 
$\sigma = 0.5 (+0.4, -0.02)$ keV with 90\% confidence.
The Fe line properties of this galaxy are more characteristic of Seyfert 1s,
where the reprocessing material is inferred to be
close to the galaxy nucleus \citep{Nan97b}, as opposed to the
narrow K$\alpha$ lines at $E = 6.4$ keV due to neutral Fe farther
from the nucleus,
which is typical of Seyfert 2s, including Seyfert 2/starburst composites
\citep{LWH01s}. 

\subsection{Other Sources\label{subsec:other}}
The group members NGC 6215 and NGC 6215A are included in the larger
field of view of the GIS detectors, and we
analyze their X-ray spectra here.
NGC 6215 lies near the extreme edge of the field of view, so we
extracted the spectrum from an elliptical region having semimajor axes
$5\farcm6$ and $3\farcm2$ to maximize the included emission.
The spectrum of NGC 6215A was extracted from a circular region
of radius $4\farcm3$.  
Source-free regions of each detector were used for the corresponding
background measurements.

With the limited data, only
simple models are appropriate.  In both cases, we fit the data
reasonably well with a single power law.  The model parameters
are listed in Table 3.
NGC 6215 has been identified as a weak LINER \citep{Dur88}.
The relatively strong X-ray emission of NGC 6215A suggests
that this galaxy also contains an active nucleus.
In addition to NGC 6215 and NGC 6215A, we observe 
the Seyfert 2 ESO 138-G1 in the
GIS fields.  \citet{Col00} discuss the spectrum
of ESO 138-G1 and find that 
either a partial covering or Compton reflection model fits it well.

The GIS3 calibration source blocks the cluster 1RXSJ165259.4-594302,
and we observe it in the GIS2 detector alone.
The cluster is centered outside the GIS field,
and we extract its spectrum from a $6\farcm7$ circular region
on the detector. 
We fit the spectrum with a thermal bremsstrahlung model 
having $kT = 2.8$ keV absorbed
by fixed Galactic column density.  The model parameters are
listed in Table 3.

\section{The Obscuring Starburst Model\label{sec:model}}

\begin{figure*}[ht]
\centering
\includegraphics[angle=270,width=6.2in]{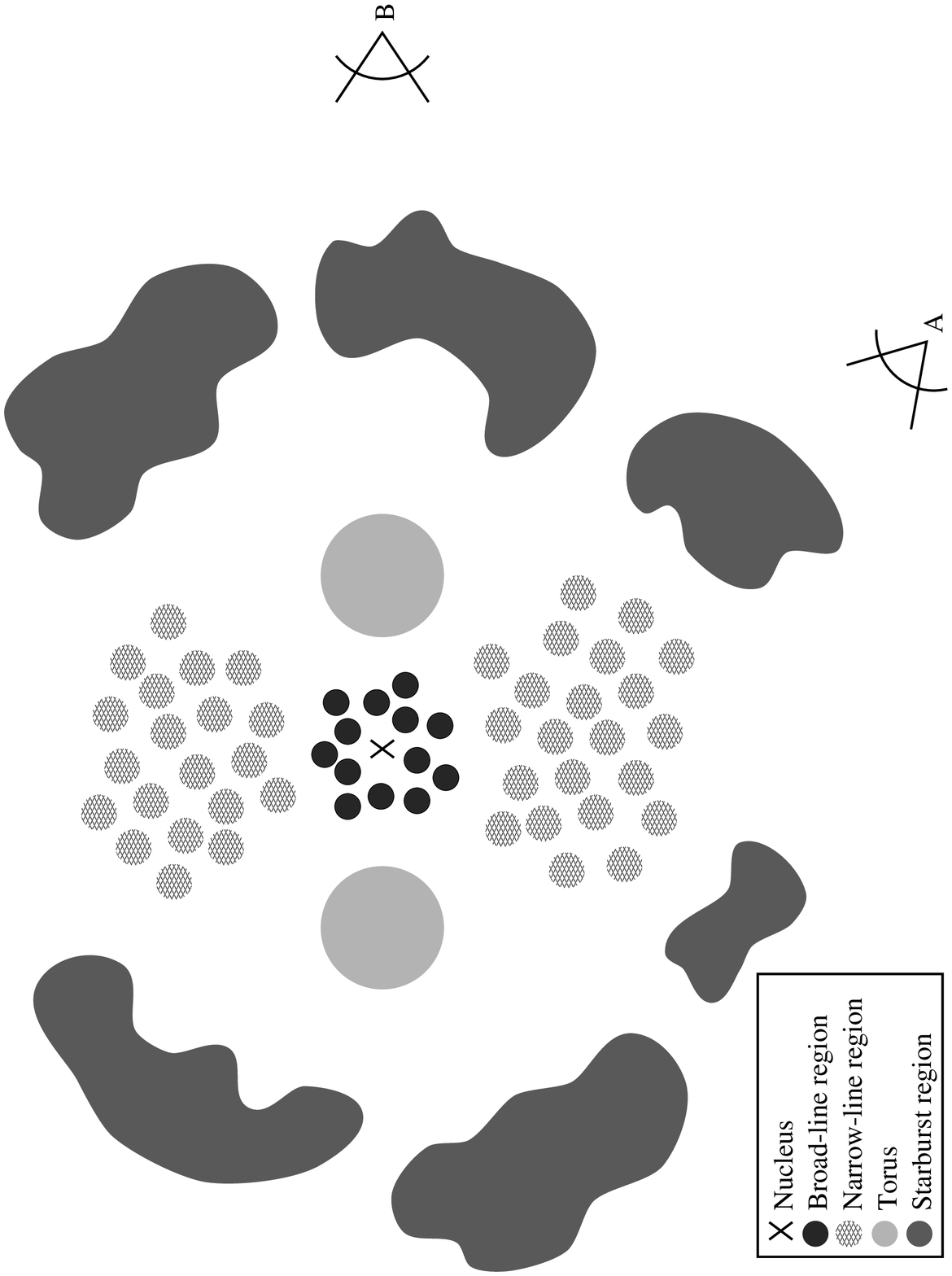}
\caption{\label{fig:cartoon}
Cartoon of the obscuring starburst model.  
We view the nucleus of NGC 6221 from position A, through
the gas and dust of the starburst.  Without the starburst,
NGC 6221 would appear to be a normal Seyfert 1.  It is not
a true Seyfert 2/starburst composite, in which 
our line of sight would pass through both the
star-forming region and the ``torus'' 
very close to the central engine (viewpoint B).
}
\end{figure*}

To account for the optical and X-ray observations of NGC 6221,
we suggest a scenario in which 
{\em the gas and dust associated with the starburst
is the material that obscures the AGN.}
The extent of the starburst-like optical
emission line ratios indicates that the physical scale of the starburst
is a few hundred parsecs.  This size is sufficient to block not only the
broad-line region, but also the narrow-line region of the AGN.  This
is unlike
the putative obscuring ``torus'' that distinguishes type 1 and type 2 Seyfert
galaxies because the latter hides only the broad-line region 
(Figure \ref{fig:cartoon}). 
The total obscuring column density to the AGN measured at 
X-ray energies is only $10^{22} \psc$.
The Seyfert 1-like characteristics of NGC 6221, such as variability
and Fe K$\alpha$ line properties, are therefore not surprising.
Without the obscuring starburst, we would have a direct view of
the AGN (Figure \ref{fig:cartoon}, viewpoint A).  
The column density to the nucleus of 
a true Seyfert 2 with an obscuring starburst is even greater
(viewpoint B), 
as both the star-forming region and the material very close
to the central engine (the ``torus'') block the view \citep{LWH01j}.

The starburst itself is dusty, as the optical spectra demonstrate.
The extinction $A_V = 3.0$ mag (\S\ref{sec:optspec}) 
is consistent with the model geometry, which
places the starburst along the line of sight to the AGN.
For $N_H = 2\times 10^{21} A_V \psc$ 
\citep{Gor75}, we have $N_H= 6\times 10^{21} \psc$.  
This measures the minimum
obscuration due to the starburst, since most of the detected
emission comes
from its exterior regions that are not so extremely absorbed.
Specifically, the direct conversion from $A_V$ to $N_H$ assumes 
that all the dust is located in a uniform
foreground screen of material, separate from
the emission sources.
As more complex model calculations demonstrate, the
effective extinction that we measure can severely underestimate
the total dust and gas content of the starburst.  In the starburst
model of \citet{Wit92}, for example, in which the stars are more
centrally concentrated than the embedding dust, the effective
extinction is only $A_V=1.7$ even when the total extinction to
the center of the galaxy reaches $A_{V, total}=15$.
Similarly, with a clumpy obscuring medium near the stars,
\citet{Cal94} find $A_V= 2$ would be measured from the Balmer decrement
when the total extinction along the line of sight $A_{V, total} = 10$,
assuming a large number of individual clumps.  

In addition to the distinct physical location of the X-ray and 
optical emission, a second uncertainty in comparing the
X-ray and optical measurements arises because
they are directly sensitive to different obscuring material.  
Photoelectric interaction with heavy elements is the immediate 
culprit at X-ray energies, and determining the hydrogen column density
then is a function of abundance.  Similarly, the optical extinction
is due to dust, and the equivalent hydrogen column density depends on
the gas-to-dust ratio.  
Nevertheless, the lower limit on column density of the starburst is
at least comparable to $N_H$ that obscures the AGN as measured in 
X-rays, and more complex geometries of emission sources and
dust within the starburst
indicate that it is likely to be greater,
which implies that the starburst intervening along the
line of sight to the active nucleus of NGC 6221 is the primary
source of obscuration.
Note that because the extinction we measure is to the 
extended starburst region, we expect standard Galactic
$A_V/N_H$ ratios to be appropriate, in contrast with
some AGNs, in which extinction to the broad-line
region is lower than expected for the absorbing column
density measured in X-rays and may be due to fundamentally different
distributions of grain size or dust composition close
to the central engine \citep[e.g.,][]{Mai00}.

The presence of both a starburst and an AGN in this geometry
naturally results in
observed [\ion{O}{3}] lines that are broader than other lines
if the two components are not resolved.
The detectable AGN contribution to Balmer lines and other features
that are characteristically strong in starbursts is relatively
small.  
The starburst is relatively weaker in [\ion{O}{3}], however, 
so the fraction
of observed luminosity in this line that is due to the AGN is
greater.  The AGN line broadening then weights the total measured 
line width more heavily. 
Broadened [\ion{O}{3}] lines in fact indicate the presence of AGNs
in substantial fractions of galaxies with starburst 
and \hii-region-like
nuclei \citep{Ken89}, in 
addition to NGC 6221 and the other X-ray loud composites.

High spatial resolution observations of
starburst galaxies demonstrate that 
the starbursts are inhomogenous \citep[e.g.,][]{Cal97}, 
so in this case, we do not expect
the AGN to be covered uniformly.  
While the physical scale of individual clumps may
be as small as distinct star-forming clouds, however, the net
obscuration to the center of NGC 6221 is large.
Similar to other starburst galaxies, in which total nuclear
column densities
of $10^{22}$ to $10^{25} {\rm \, cm^{-2}}$ are typical
(e.g., \citealt*{Dah98}; \citealt{Ken98rev}),
the line of sight intersects many dusty regions, so the AGN is
never directly exposed.
If changes in the intrinsic
AGN and its obscuring medium can be distinguished, X-ray variability 
measurements may best constrain the physical scale of the obscuring
clouds.

\begin{center}
\includegraphics[width=3.5in]{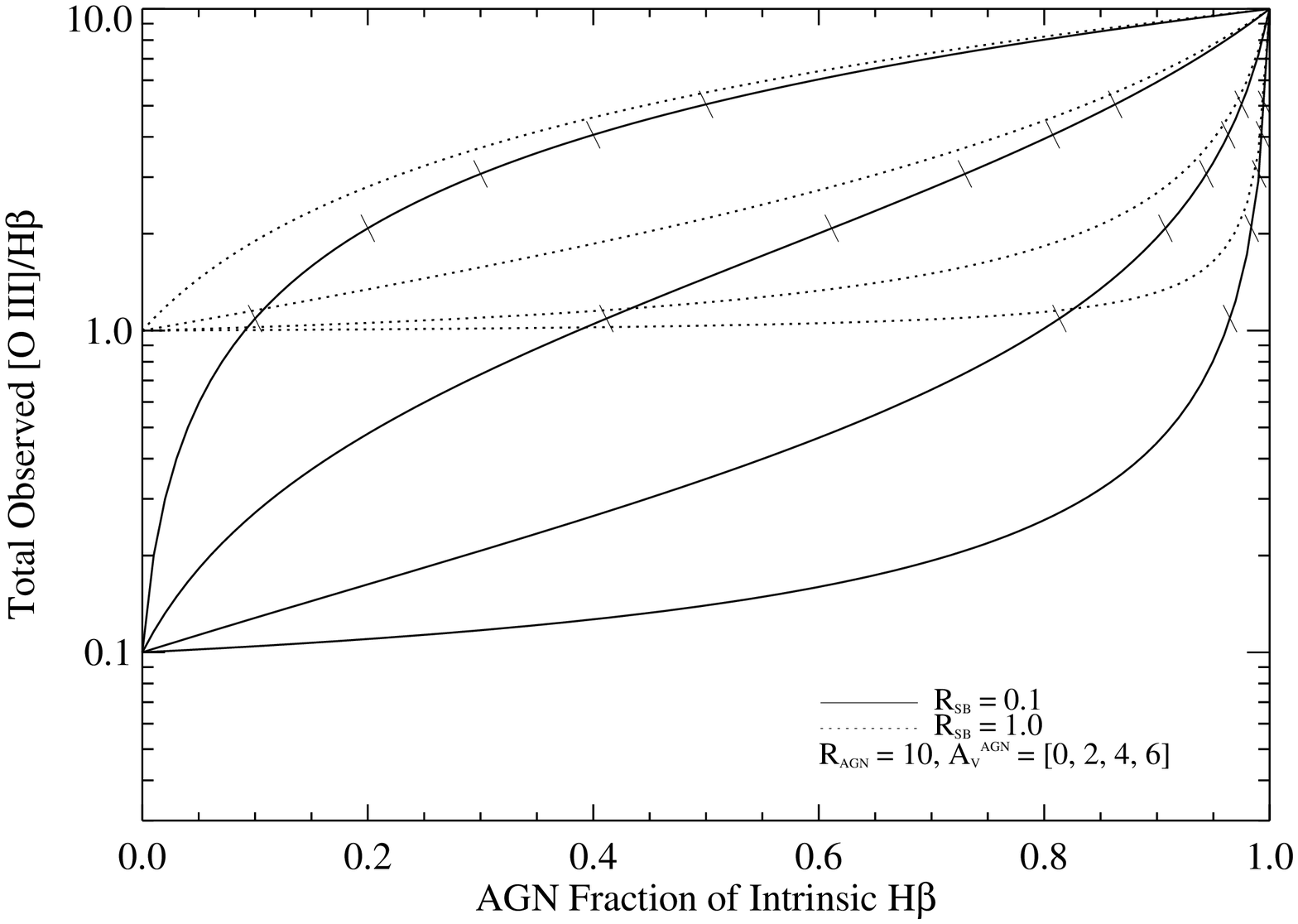}
\end{center}
\vskip -0.1in
\figcaption{\label{fig:rvarysb}
Net observed ratio, $R$, of [\ion{O}{3}]/H$\beta$ intensity vs. AGN fraction of
total intrinsic H$\beta$ flux, calculated for $R_{AGN}=10$,
$A_V^{AGN} = 0$, 2, 4, and 6 (increasing from upper left to lower right), 
and $R_{SB}=0.1$ and 1.0.  
With some obscuration, even a strong
AGN may not be detected in the observed line ratios.
The AGN fraction of {\em observed} H$\beta$ flux
is marked with ticks corresponding to 
0.1, 0.2, 0.3, 0.4, and 0.5 along the $R_{SB}=0.1$ curves.
The relationship between  the observed and intrinsic 
H$\beta$ fractions depends only on $A_V^{AGN}$, not $R_{SB}$ or $R_{AGN}$.
For example, with $A_V^{AGN} = 2$, the observed H$\beta$ fraction equals
0.1 when the intrinsic H$\beta$ fraction is 0.4.
}
\vskip 0.1in

The \hst{} images demonstrate that the observed optical flux due to the
AGN is small compared with the 
surrounding starburst.
Both sources are strongly extincted, and
the AGN is also {\em intrinsically} 
weak with respect to the starburst at optical wavelengths.
We roughly estimate the AGN properties from the X-ray luminosity.  
First we estimate the 
broad H$\beta$ line luminosity from the 
hard X-ray luminosity, $L_{2-10}$.
\citet*{Xu99} compile the X-ray measurements, and we find the corresponding
H$\beta$ values of the Seyfert 1s in \citet{Whi92} and \citet{Dah88}.  
The median ratio of these is $L_{H\beta}^{BLR}/L_{2-10} = 0.013$.
In the high-flux state of NGC 6221, as observed with \asca, 
$L_{2-10}= 6.3 \times 10^{41} {\rm \,erg\,s^{-1}}$, 
so we predict
$L_{H\beta}^{BLR} = 8.5\times 10^{39}{\rm \,erg\,s^{-1}}$, or
flux 
$F_{H\beta}^{BLR} = 1.8\times10^{-13} {\rm \,erg\,cm^{-2}\,s^{-1}}$
from the {\em intrinsic} AGN.
Expecting 
$A_V = 5.0$
to the broad-line region based on the 
X-ray column density, the resulting expected {\em detectable} flux 
$F_{H\beta}^{BLR}=1.8 \times10^{-15} {\rm \,erg\,cm^{-2}\,s^{-1}}$,
which is well below the upper limit on detection of the
BLR estimated from the optical spectrum.
Even adopting the minimum $A_V = 3.0$ to the AGN
based on  the optical data, 
the expected measured flux
$F_{H\beta}^{BLR}=1.1 \times10^{-14} {\rm \,erg\,cm^{-2}\,s^{-1}}$,
at the detection limit in the optical spectrum
and less than 
the total observed $F_{H\beta}=2.4\times10^{-14} {\rm \,erg\,cm^{-2}\,s^{-1}}$
in the nuclear region.
Using the X-ray--Balmer line luminosity correlation of \citet{Ho01}
for broad-line galaxies produces similar results, with the weak BLR
undectable in our optical spectrum.

\begin{center}
\includegraphics[width=3.5in]{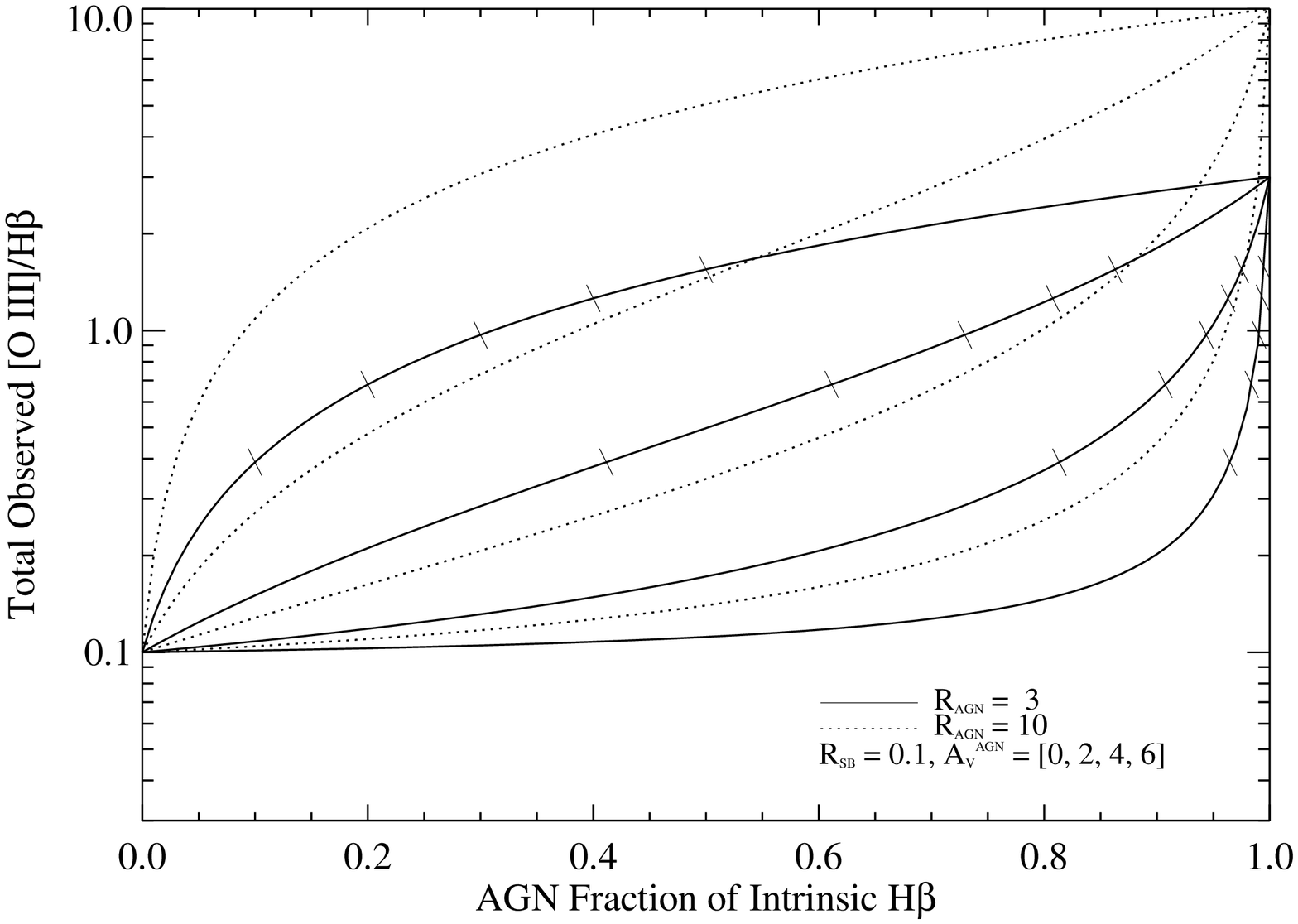}
\end{center}
\vskip -0.1in
\figcaption{\label{fig:rvaryagn}
Net observed ratio, $R$, of [\ion{O}{3}]/H$\beta$ intensity vs. AGN fraction of
total intrinsic H$\beta$ flux, calculated for $R_{SB}=0.1$,
$A_V^{AGN} = 0$, 2, 4, and 6 (increasing from upper left to lower right), 
and $R_{AGN}=3$ and 10.  
The AGN fraction of {\em observed} H$\beta$ flux
is marked with ticks corresponding to 
0.1, 0.2, 0.3, 0.4, and 0.5 along the $R_{AGN}=3$ curves.
The combination of obscuration and relatively low AGN power can produce
line ratios typical of pure starbursts.
}
\vskip 0.1in

Second, we crudely estimate the intrinsic AGN 
[\ion{O}{3}]$\lambda 5007$ luminosity, 
which is produced in the NLR, 
from $L_{2-10}$.  
The median ratio of these two quantities 
$L_{[O III]}^{NLR}/ L_{2-10} = 8.7\times 10^{-3}$ 
for the Seyfert 1s of the \citet{Xu99}
compilation.
For the observed $L_{2-10} = 6.6\times10^{41} {\rm \,erg\,cm^{-2}\,s^{-1}}$, 
$L_{[O III]}^{NLR}=5.7\times10^{39}{\rm \,erg\,cm^{-2}\,s^{-1}}$,  
or intrinsic flux $F_{[O III]}^{NLR} = 1.2\times 10^{-13} {\rm \,erg\,cm^{-2}\,s^{-1}}$.
The advantage of calibrating this relationship with only the Seyfert 1s is that 
the ratio $L_{[O III]}^{NLR}/L_{2-10}$ 
is sensitive to intervening column density,
although fitting all radio quiet galaxies in the data set produces similar
results.
From the AGN component observed at optical
wavelengths, we expect the reddening-corrected 
$F_{[O III]}^{NLR} = 4.2\times 10^{-13} {\rm \,erg\,cm^{-2}\,s^{-1}}$
in the $1\farcs5 \times 1\farcs5$ central aperture,
or $F_{[O III]}^{NLR} = 8.1\times 10^{-13} {\rm \,erg\,cm^{-2}\,s^{-1}}$,
integrated over the whole slit.
Two uncertainties relevant to both of these
X-ray estimates are that the data set of X-ray and
optical fluxes is not uniform
and the optical fluxes of this sample have not been corrected for reddening.
The scatter in the data is large, which may therefore account for
our underprediction of the observed \o3{} luminosity.
Although the flux measured in the optical spectrum 
is approximately 
four times greater than the intrinsic \o3{} flux we predict based
on the hard X-ray flux, our main point is to demonstrate that
the X-ray and optical data all provide a generally consistent
description of the starburst and AGN in NGC 6221.

\begin{center}
\includegraphics[width=3.5in]{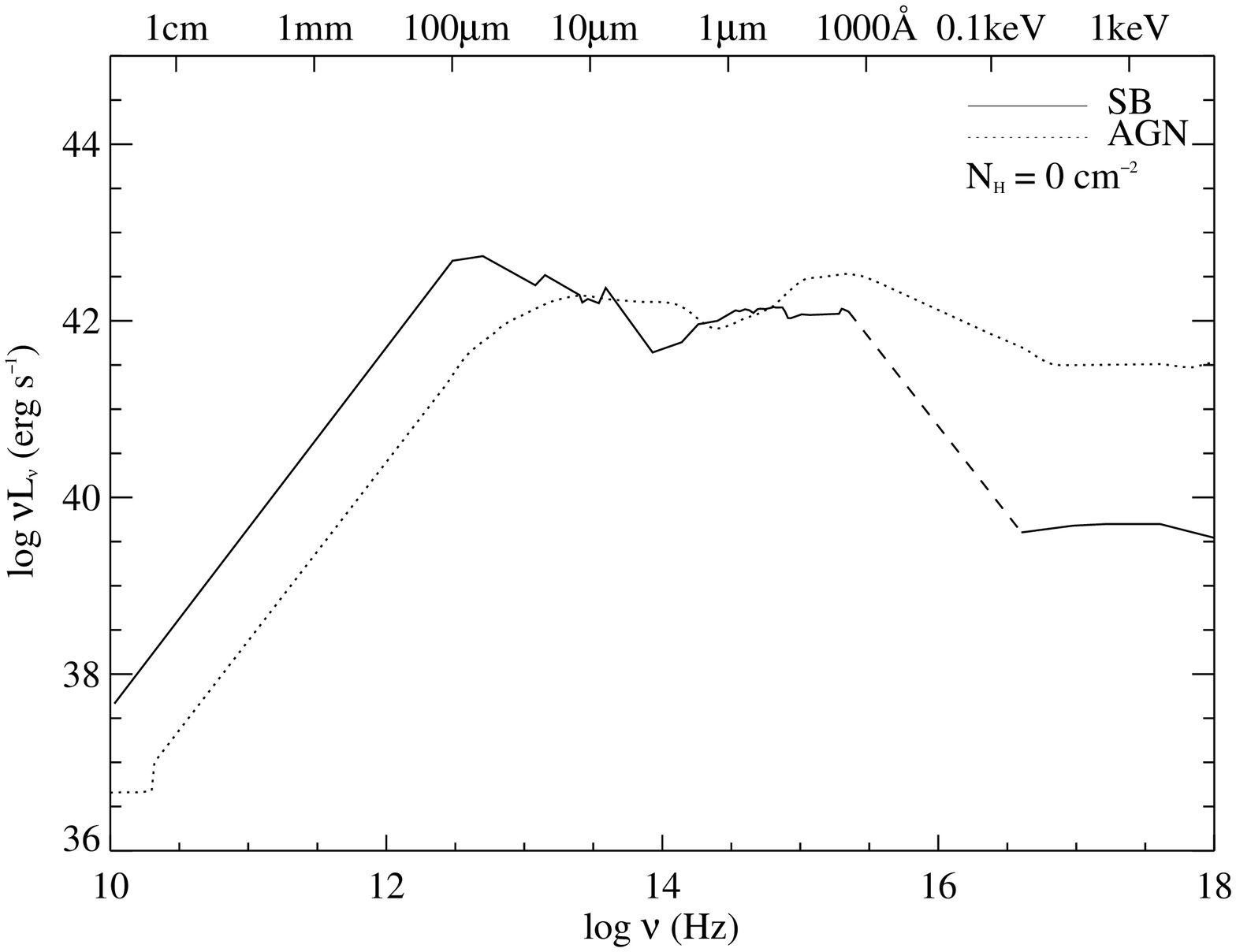}
\end{center}
\vskip -0.1in
\figcaption{\label{fig:sed}
Spectral energy distributions of unobscured 
sources illustrate the 
X-ray prominence of AGNs.  The 
average intrinsic starburst ({\it solid line}; Schmitt et al. 1997) 
and radio-quiet quasar ({\it dotted line}; Elvis et al. 1994),
are each normalized at 5000\AA. 
We model the X-ray spectrum of the starburst as
thermal emission due to gas at $kT = 1$ keV, and the dashed
line covers the UV region where there intrinsic spectrum is
unknown.
When the optical fluxes are equal, the AGN is two orders of
magnitude brighter than the starburst in X-rays, although 
the bolometric luminosities
of the different emission sources are comparable.
}
\vskip 0.1in

Mixing starburst and AGN components can produce an ambiguous
optical spectrum, as \citet{Hil01} and \citet{Nel01} demonstrate
for several diagnostic line ratios.  
Additional obscuration of the AGN can completely hide this
component to produce an optical spectrum that mimics a pure starburst.
We compute some simple models to illustrate the detected changes
in the important flux ratio $R\equiv F_{[O III]}/F_{H\beta}$
in general circumstances.  
In the first
example (Figure \ref{fig:rvarysb}), we adopt a typical AGN ratio
$R_{AGN}=10$ and  
optical extinction to the starburst $A_V^{SB} = 0$, and  we
consider the effects of additional extinction 
to the AGN ($A_V^{AGN} = 0$, 2, 4, and 6)  
and different starburst line ratios 
($R_{SB} = 0.1$, and 1.0). 
Even when the 
intrinsic AGN is strong, providing up to 50\% of the completely 
unobscured narrow H$\beta$ luminosity, the net observed $R$ can be
indistinguishable from a pure starburst.  
Similarly, we 
consider $R_{AGN}=3$ and 10 and show that 
for the typical $R_{SB}=0.1$, some combination of obscuration and 
relative AGN luminosity can hide the AGN in the observed line
ratio (Figure \ref{fig:rvaryagn}).
In all cases, only the additional obscuration of the AGN, $A_V^{AGN}-A_V^{SB}$,
is relevant, and the plots are identical for $A_V^{SB}=3$ and
$A_V^{AGN} = 3$, 5, 7, and 9, which encompasses the likely
parameters in the case of NGC 6221.
We also note the observed fraction of H$\beta$ flux due
to the AGN with ticks corresonding to observed fractions
0.1, 0.2, 0.3, 0.4, and 0.5.  For example, with $A_V^{AGN} = 2$,
10\% of the detected H$\beta$ emission is due to the AGN when
its intrinsic fraction is 0.4, while for $A_V^{AGN}=4$, 
10\% of the detected H$\beta$ emission is due to the AGN only
when the 
an intrinsic AGN fraction reaches 0.8.
The relationship between the intrinsic and observed H$\beta$
fractions depends only on $A_V^{AGN}$, not $R_{SB}$ or $R_{AGN}$.

The effect of obscuration on typical spectral energy distributions also
illustrates that an AGN can be hidden at optical wavelengths while
remaining the dominant X-ray source in composite galaxies.
We compare the average unobscured starburst galaxy 
\citep{Schm97} with the median radio-quiet QSO \citep{Elv94},
normalized at 5000\AA{} (Figure \ref{fig:sed}).  
The bolometric luminosities of the two sources are comparable,
with the starburst much stronger in the FIR and the AGN
dominating in X-rays.

\begin{center}
\includegraphics[width=3.5in]{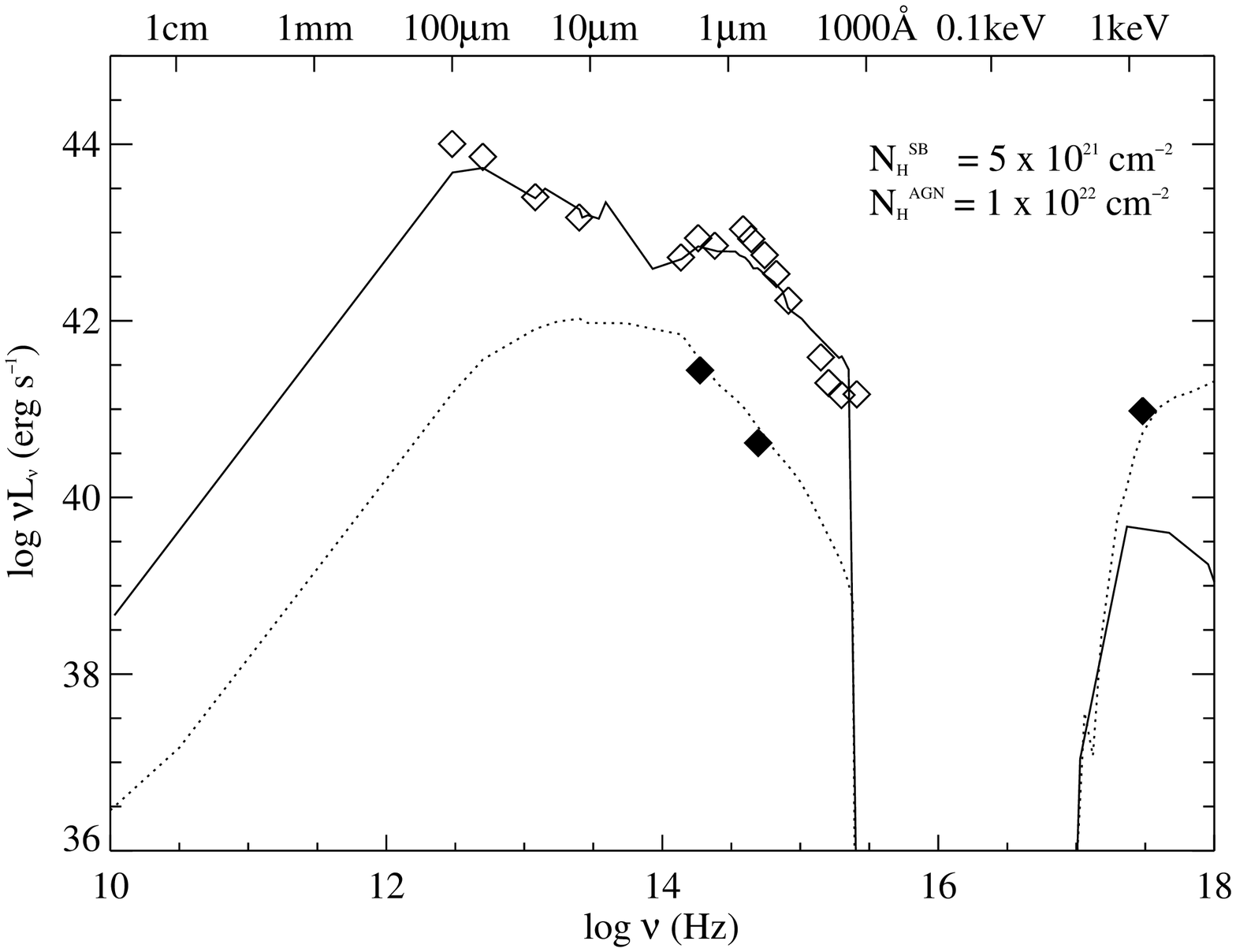}
\end{center}
\vskip -0.1in
\figcaption{\label{fig:n6sed}
Observed spectral energy distribution of NGC 6221, modelled
with typical AGN and starburst contributions. 
The AGN component is measured in X-rays and
as the point source alone in \hst{} images ({\it filled symbols}).
Imaging and spectroscopy of NGC 6221 that encompass larger scales
are predominantly sensitive to the starburst ({\it open symbols}).
The average AGN ({\it dotted line}) 
is extincted by $1\times 10^{22} {\rm \, cm^{-2}}$,
as we measure in the X-ray spectrum of NGC 6221, 
while the average starburst ({\it solid line}) 
is extincted by $5\times 10^{21} {\rm \, cm^{-2}}$, as 
we measure optically.
In NGC 6221, the AGN is intrinsically weak compared
with the starburst, as well as heavily absorbed.
While the starburst accounts for nearly all  the bolometric
luminosity of NGC 6221, the AGN still produces its X-ray
emission.
Additional data in the FIR, optical--NIR, and UV bands
are from 
\citet{San95}, 
\citet{McA83}, and
\citet{Kin93}, respectively.
}
\vskip 0.1in

We model NGC 6221 with these two components, 
scaling the empirical spectral energy
distributions to the data values and applying
the measured extinctions (Figure \ref{fig:n6sed}).
The X-ray and \hst{} measurements of the point source indicate
the contribution of the AGN, and we use the X-ray determination of
its extinction.  Lower-resolution imaging and large-aperture
spectroscopy measure the starburst component, and the
Balmer decrement indicates its exctinction.  
In this calculation, we adopt the extinction laws of
\citet{Car89} at long wavelengths, that of \citet{Cal00} in the NIR-to-UV
bandpass, and that of \citet{Mor83} at higher energies.
While the empirical \citet{Cal00} law is based on observations
of starburst galaxies, it is also reasonably applied 
to the AGN component.  Specifically, it is relatively grey and
lacks the strong Galactic feature at 2175\AA{}, which is not 
detected in distant quasars \citep[and references therein]{Pit00}. 
The AGN is much weaker than the starburst
at all wavelengths except X-rays. 
The AGN accounts for only about 1\% of the galaxy's bolometric
luminosity, while providing nearly all the observed
X-rays. 
In NGC 6221 and other X-ray loud composites,
the dusty starburst is both the obscuring medium 
and the origin of the detectable optical and infrared spectrum.

The population of starburst-obscured AGNs helps to solve the
problem of the spectrum of the XRB.
The observed XRB spectrum is much flatter than the spectra
of unobscured AGNs or any other extragalactic populations
that likely comprise the background.  A combination of
absorption and redshift of AGNs can produce the observed
XRB, however \citep{Set89,Mad94,Com95}.
As \citet{Fab98} have proposed, starburst-obscured AGNs
specifically may contribute significantly to the XRB, 
naturally providing the required 
large optical depths and high covering fractions 
of their central engines.

The starburst and AGN phenomena may be related in an evolutionary
sequence, with galactic interactions triggering the starbursts
and providing a central reservoir of
material for accretion in the active nucleus \citep{Cid01,Sto01}.
The limited lifetime of the starburst then constrains the
timescale of the strongly obscured AGN phase.  Typical starbursts
persist only over timescales up to several hundred Myr.
As a result of the starburst, 
material is blown out
of the galaxy by the superwind or converted into dense
stellar remnants, and no longer blocks the active
nucleus.  
Thus, as \citet{Hec89} describe, only in a later phase of evolution are
direct views of the central engine possible,
with characteristic optical spectra that include
broad emission lines.

\section{Summary and Significance}
We have presented NGC 6221 as a case study of an X-ray loud composite
galaxy.
While the 
optical spectrum of the nuclear region
is typical of a starburst galaxy, its buried AGN is clearly
revealed in X-rays.  The Balmer decrement indicates that the
starburst itself is strongly reddened, and the column density
to the AGN measured in X-rays is even greater.  We suggest that
the very red point source distinguished in \hst{} images is the
nucleus, and we crudely derive intrinsic optical properties of
the AGN from the X-ray luminosity.
We demonstrate generally
that the combination of obscuration and contrast 
in viewing an intrinsically weak AGN against a bright starburst
can produce the net optical spectrum of the surrounding
starburst, while the active nucleus remains the dominant X-ray source.

Without the obscuring starburst, NGC 6221 would be classified
as a Seyfert 1, with the central engine and broad-line
region viewed directly.
In the simple physical model, the starburst need not completely
block the AGN from all lines of sight, but its covering fraction may
be larger than that of the dense material very close to
the nucleus---the ``torus''---that accounts for some of 
the observed distinctions between Seyfert 1s and 2s.
In some cases, the starburst may 
block only the broad emission lines, not the narrow ones.
Consequently, if such obscuring starbursts are common,
the observed ratios of Seyfert 1s and 2s do not directly 
constrain the ``torus''  characteristics such
as covering fraction.

NGC 6221 is not unique, and such objects probably occur more frequently
than their small fraction (7/210) in the joint {\it ROSAT-IRAS}
sample \citep{Bol92,Mor96} would suggest.  Because the \rosat{}
survey is sensitive only to soft (0.2--2 keV) X-rays, it misses
the more heavily obscured members of this class.
A column density of $10^{22} {\rm \, cm^{-2}}$ reduces the
flux in the \rosat{} bandpass by a factor of 10, and
$N_H= 10^{23} {\rm \, cm^{-2}}$ reduces the flux by more than three
orders of magnitude.  These column densities are not extreme
compared with the observed
nuclear column densities of molecular gas in 
ordinary starburst galaxies.
While optical spectroscopy is the primary method of identifying
the sources detected through efforts to resolve the X-ray
background, it does not always reveal the complex nature of
these galaxies.
Inverting the problem---considering the X-ray emission of
known optically-identified 
populations---also overlooks the contribution of the X-ray loud 
composite AGNs.
Hard X-ray surveys are the best method to detect this class of
galaxies, and they require careful spectroscopy at lower
energies to correctly identify 
their intrinsic emission sources.
Future missions with greater sensitivity to harder X-rays
will prove most effective at finding X-ray loud AGNs. 
{\it Constellation-X}, for example,
is expected to have an effective area of about 5000 cm$^2$ at
10 keV and 1500 cm$^2$ at 40 keV, and will therefore sensitively
detect obscured AGNs of all kinds.

\acknowledgements
We thank Henrique Schmitt for many valuable conversations and his 
lively interest in the subject.
This work was supported by NASA grants  NAG5-6917 and NAG5-6400.
RCF was supported by the National Science Foundation through 
grant GF-1001-99 from the Association of Universities for Research in
Astronomy, Inc., under NSF cooperative agreement AST-9613615. 
This work is  based on observations made with the NASA/ESA 
{\it Hubble Space Telescope}, obtained from the data archive at 
the Space Telescope Science Institute.
STScI is operated by the Association of Universities for Research 
in Astronomy, Inc. under NASA contract NAS 5-26555.
This research has also made use of the NASA/IPAC Extragalactic Database
(NED) which is operated by the Jet Propulsion Laboratory,
California Institute of Technology, under contract with the
National Aeronautics and Space Administration,
the Astronomical Data Center at NASA Goddard Space Flight Center,
and the High Energy Astrophysics Science Archive Research Center Online Service
provided by the NASA Goddard Space Flight Center.


\begin{thebibliography}{}
\bibitem[Binette, Fosbury, \& Parker(1993)]{Bin93}Binette, L., Fosbury, R.\ A., \& Parker, D.\ 1993, \pasp, 105, 1150 
\bibitem[Boller \ea (1992)]{Bol92} Boller, T., Meurs, E. J. A., Brinkmann, W., Fink, H., Zimmermann, U., \& Adorf, H.-M. 1992, \aap, 261, 57 
\bibitem[Boroson \& Green(1992)]{Bor92}Boroson, T.\ A., \& Green, R.\ F.\ 1992, \apjs, 80, 109 
\bibitem[Calzetti et al.(2000)]{Cal00}Calzetti, D., Armus, 
L., Bohlin, R.\ C., Kinney, A.\ L., Koornneef, J., \& Storchi-Bergmann, T.\ 
2000, \apj, 533, 682 
\bibitem[Calzetti, Kinney, \& Storchi-Bergmann(1994)]{Cal94}Calzetti, D., Kinney, A.\ L., \& Storchi-Bergmann, T.\ 1994, \apj, 429, 582 
\bibitem[Calzetti et al.(1997)]{Cal97}Calzetti, D., Meurer, G.\ R., Bohlin, R.\ C., Garnett, D.\ R., Kinney, A.\ L., Leitherer, C., \& Storchi-Bergmann, T.\ 1997, \aj, 114, 1834 
\bibitem[Cardelli \ea(1989)Cardelli, Clayton \& Mathis]{Car89}Cardelli, 
J.\ A., Clayton, G.\ C., \& Mathis, J.\ S.\ 1989, \apj, 345, 245 
\bibitem[Chevalier \& Clegg(1985)]{Che85}Chevalier, R. A., \& Clegg, A. W. 1985, Nature, 317, 44
\bibitem[Cid Fernandes \ea(2001)]{Cid01}Cid Fernandes, R., Jr., Heckman, T. M., Schmitt, H., Gonz\'alez Delgado, R. M., \& Storchi Bergmann, T. 2001, submitted to \apj
\bibitem[Cid Fernandes \ea(1998)Cid Fernandes, Storchi-Bergmann, \& Schmitt]{Cid98}Cid Fernandes, R., Jr., Storchi-Bergmann, T., \& Schmitt, H. R. 1998, \mnras, 297, 579
\bibitem[Collinger \& Brandt(2000)]{Col00}Collinger, M. J., \& Brandt, W. N. 2000, \mnras, 317, L35
\bibitem[Comastri et al.(1995)]{Com95}Comastri, A., Setti, G., Zamorani, G., \& Hasinger, G.\ 1995, \aap, 296, 1 
\bibitem[Dahari \& de Robertis(1988)]{Dah88}Dahari, O., \& de Robertis, M. M. 1988, \apjs, 67, 249 
\bibitem[Dahlem \ea (1998)Dahlem, Weaver, \& Heckman]{Dah98}Dahlem, M., Weaver, K. A., \& Heckman, T. M. 1998, \apjs, 118, 401
\bibitem[Durret \& Bergeron(1988)]{Dur88}Durret, F., \& Bergeron, J. 1988, \aaps, 75, 273 
\bibitem[Elvis \ea (1994)]{Elv94}Elvis, M., et al. 1994, \apjs, 95, 1 
\bibitem[Fabian \ea (1998)]{Fab98}Fabian, A.\ C., Barcons, X., Almaini, O., \& Iwasawa, K.\ 1998, \mnras, 297, L11 
\bibitem[Giacconi \ea (1962)]{Gia62}Giacconi, R., Gursky, H., Paolini, F. R., \& Rossi, B. B. 1962, Phys. Rev. Letters 9, 439
\bibitem[Giacconi \ea (2001)]{Gia01}Giacconi, R., et al. 2001, \apj, in press (astro-ph/0007240) 
\bibitem[Gonz{\'a}lez Delgado \ea(2001)Gonz{\'a}lez Delgado, Heckman, \& Leitherer]{Gon01}Gonz{\'a}lez Delgado, R. M., Heckman, T. M., \& Leitherer, C. 2001, \apj, 546, 845
\bibitem[Gonz{\'a}lez Delgado et al.(1998)]{Gon98}Gonz{\'a}lez Delgado, R. M., Heckman, T., Leitherer, C., Meurer, G., Krolik, J., Wilson, A. S., Kinney, A., \& Koratkar, A. 1998, \apj, 505, 174 
\bibitem[Gorenstein(1975)]{Gor75}Gorenstein, P. 1975, \apj, 198, 95 
\bibitem[Hasinger \ea (1998)]{Has98} Hasinger, G., Burg, R., Giacconi, R., Schmidt, M., Trumper, J., \& Zamorani, G. 1998, \aap, 329, 482 
\bibitem[Heckman \ea(1997)]{Hec97}Heckman, T. M., et al. 1997, \apj, 482, 114 
\bibitem[Heckman et al.(1989)]{Hec89}Heckman, T. M., Blitz, L., Wilson, A. S., Armus, L., \& Miley, G. K. 1989, \apj, 342, 735
\bibitem[Heckman, Lehnert, \& Armus(1993)]{Hec93}Heckman, T., Lehnert, M., \& Armus, L. 1993, in ``The Environment and Evolution of Galaxies,'' ed. Shull \& Thronson (Kluwer: Dordrecht), 455
\bibitem[Hill \ea(2001)]{Hil01}Hill, T. L., Heisler, C. A., Norris, R. P., Reynolds, J. E., \& Hunstead, R. W. 2001, \aj, 212, 128
\bibitem[Ho et al.(2001)]{Ho01}Ho, L.\ C., et al.\ 2001, \apjl, 549, L51 
\bibitem[Hornschemeier \ea (2000)]{Hor00}Hornschemeier, A. E., et al. 2000, \apj, 541, 49 
\bibitem[Kennicutt(1998a)]{Ken98sfr}Kennicutt, R. C., Jr. 1998a, \apj, 541, 552
\bibitem[Kennicutt(1998b)]{Ken98rev}Kennicutt, R. C., Jr. 1998b, \araa, 36, 189
\bibitem[Kennicutt, Keel, \& Blaha(1989)]{Ken89}Kennicutt, R.\ C., Keel, W.\ C., \& Blaha, C.\ A.\ 1989, \aj, 97, 1022 
\bibitem[Kinney et al.(1993)]{Kin93}Kinney, A.\ L., Bohlin, R.\ C., Calzetti, D., Panagia, N., \& Wyse, R.\ F.\ G.\ 1993, \apjs, 86, 5 
\bibitem[Lehnert \& Heckman(1996)]{Leh96}Lehnert, M. D., \& Heckman, T. M. 1996, \apj, 462, 651 
\bibitem[Leitherer \ea(1999)]{Lei99}Leitherer, C., et al.\ 1999, \apjs, 123, 3 
\bibitem[Levenson \ea(2001a)Levenson, Weaver, \& Heckman]{LWH01j}Levenson, N. A., Weaver, K. A., \& Heckman, T. M. 2001a, \apj, 549, in press
\bibitem[Levenson \ea(2001b)Levenson, Weaver, \& Heckman]{LWH01s}Levenson, N. A., Weaver, K. A., \& Heckman, T. M. 2001b, \apjs, 133, in press
\bibitem[Madau, Ghisellini, \& Fabian(1994)]{Mad94}Madau, P., Ghisellini, G., \& Fabian, A.\ C.\ 1994, \mnras, 270, L17 
\bibitem[Maiolino \ea (2001)]{Mai00}Maiolino, R., Marconi, A., Salvati, M., Risaliti, G., Severgnini, P., Oliva, E., La Franca, F., \& Vanzi, L. 2001, \aap, 365, 28
\bibitem[Marshall et al.(1979)]{Mar79}Marshall, F.\ E., Boldt, E.\ A., Holt, S.\ S., Mushotzky, R.\ F., Rothschild, R.\ E., Serlemitsos, P.\ J., \& Pravdo, S.\ H.\ 1979, \apjs, 40, 657 
\bibitem[McAlary et al.(1983)]{McA83}McAlary, C.\ W., McLaren, R.\ A., McGonegal, R. J., \& Maza, J.\ 1983, \apjs, 52, 341 
\bibitem[Meurer, Heckman, \& Calzetti(1999)]{Meu99} Meurer, G.\ R., Heckman, T.\ M., \& Calzetti, D.\ 1999, \apj, 521, 64 
\bibitem[Moorwood \& Oliva(1988)]{Moo88}Moorwood, A.\ F.\ M., \& Oliva, E.\ 1988, \aap, 203, 278 
\bibitem[Moran \ea(1996)Moran, Halpern, \& Helfand]{Mor96}Moran, E. C., Halpern, J. P., \& Helfand, D. J. 1996, \apjs, 106, 341
\bibitem[Morris \& Ward(1988)]{Mor88}Morris, S.\ L., \& Ward, M.\ J.\ 1988, \mnras, 230, 639 
\bibitem[Morrison \& McCammon(1983)]{Mor83}Morrison, R., \& McCammon, D.\ 1983, \apj, 270, 119 
\bibitem[Mushotzky \ea (2000)]{Mus00}Mushotzky, R. F., Cowie, L. L., Barger, A. J., \& Arnaud, K. A. 2000, Nature, 404, 459
\bibitem[Nandra \ea (1997a)]{Nan97a}Nandra, K., George, I. M., Mushotzky, R. F., Turner, T. J., \& Yaqoob, T. 1997a, \apj, 476, 70 
\bibitem[Nandra et al.(1997b)]{Nan97b} Nandra, K., George, I. M., Mushotzky, R. F., Turner, T. J., \& Yaqoob, T. 1997b, \apj, 477, 602 
\bibitem[Nelson \& Whittle(2001)]{Nel01}Nelson, C., \& Whittle, M. 2001, submitted to \apj
\bibitem[Osterbrock(1989)]{Ost89}Osterbrock, D. E. 1989, Astrophysics of Gaseous Nebulae and Active Galactic Nuclei (Mill Valley, CA: University Science Books)
\bibitem[Pence \& Blackman(1984)]{Pen84}Pence, W. D., \& Blackman, C. P. 1984, \mnras, 207, 9
\bibitem[Phillips(1979)]{Phi79}Phillips, M. M. 1979, \apjl, 227, L121 
\bibitem[Pitman, Clayton, \& Gordon(2000)]{Pit00}Pitman, K.\ M., Clayton, G.\ C., \& Gordon, K.\ D.\ 2000, \pasp, 112, 537 
\bibitem[Sanders et al.(1995)]{San95}Sanders, D.\ B., Egami, 
E., Lipari, S., Mirabel, I.\ F., \& Soifer, B.\ T.\ 1995, \aj, 110, 1993 
\bibitem[Schlegel, Finkbeiner, \& Davis(1998)]{Schl98}Schlegel, D.\ J., Finkbeiner, D.\ P., \& Davis, M.\ 1998, \apj, 500, 525 
\bibitem[Schmidt \ea (1998)]{Schm98}Schmidt, M., et al. 1998, \aap, 329, 495 
\bibitem[Schmitt \& Kinney(1996)]{Schm96}Schmitt, H.\ R., \& Kinney, A.\ L.\ 1996, \apj, 463, 498 
\bibitem[Schmitt \ea (1997)]{Schm97}Schmitt, H. R., Kinney, A. L., Calzetti, D., \& Storchi Bergmann, T. 1997, \aj, 114, 592 
\bibitem[Setti \& Woltjer(1989)]{Set89}Setti, G., \& Woltjer, L.\ 1989, \aap, 224, L21 
\bibitem[Storchi-Bergmann \ea(1994)Storchi-Bergmann, Calzetti, \& Kinney]{Sto94}Storchi-Bergmann, T., Calzetti, D., \& Kinney, A. L. 1994, \apj, 429, 572 
\bibitem[Storchi-Bergmann \ea(2001)]{Sto01}Storchi-Bergmann, T., Gonz\'alez Delgado, R. M., Schmitt, H., Cid Fernandes, R., \& Heckman, T. 2001, submitted to \apj
\bibitem[Storchi-Bergmann \ea(1995)Storchi-Bergmann, Kinney, \& Challis]{Sto95}Storchi-Bergmann, T., Kinney, A. L., \& Challis, P. 1995, \apjs, 98, 103 
\bibitem[Turner \ea (1999)]{Tur99}Turner, T. J., George, I. M., Nandra, K., \& Turcan, D. 1999, \apj, 524, 667 
\bibitem[Veilleux \& Osterbrock(1987)]{Vei87}Veilleux, S., \& Osterbrock, D. E. 1987, \apjs, 63, 295 
\bibitem[V\'eron \ea(1981)V\'eron, V\'eron, \& Zuiderwijk]{Ver81}V\'eron, M. P., V\'eron, P., \& Zuiderwijk, E. J. 1981, \aap, 98, 34
\bibitem[Voges \ea (1999)]{Vog99}Voges, W., et al. 1999, \aap, 349, 389 
\bibitem[Whittle(1992)]{Whi92}Whittle, M. 1992, \apjs, 79, 49
\bibitem[Witt \ea(1992)Witt, Thronson, \& Capuano]{Wit92}Witt, A.\ N., Thronson, H.\ A., \& Capuano, J.\ M.\ 1992, \apj, 393, 611 
\bibitem[Xu \ea(1999)Xu, Livio, \& Baum]{Xu99}Xu, C., Livio, M., \& Baum, S. 1999, \aj, 118, 1169
\end{thebibliography}
\end{document}